\documentclass[12pt,a4paper]{article}

\usepackage{hyperref}
\usepackage{amsmath}
\usepackage{amssymb}
\usepackage{mathrsfs}
\usepackage{subfigure}
\usepackage{graphicx}
\usepackage{color}
\usepackage{pstricks,pst-coil,pst-plot}
\usepackage[normalem]{ulem}

\numberwithin{equation}{section}

\newrgbcolor{darkred}{0.7 0.13 0.13}

\newrgbcolor{darkred}{0.7 0.13 0.13}
\newrgbcolor{darkgreen}{0 0.5 0}

\newcommand{\T}{\mathcal{T}}
\newcommand{\eps}{\epsilon}

\newcommand{\be}{\begin{equation}}
\newcommand{\ee}{\end{equation}}

\begin{document}

\title{AdS flowing black funnels: \\ Stationary AdS black holes with non-Killing horizons and heat transport in the dual CFT}

\author{Sebastian Fischetti${}^1$,
  Donald Marolf${}^{1,2}$, and Jorge E. Santos${}^1$\\
  \\
  {\it ${}^1$Department of Physics, University of California at Santa Barbara,}\\
    {\it Santa Barbara, CA 93106, U.S.A.} \\
  {\it ${}^2$Department of Physics, University of Colorado,}\\
    {\it Boulder, CO 80309, U.S.A.} \\
  }

\date{\today}

\maketitle

\begin{abstract}
We construct stationary non-equilibrium black funnels locally asymptotic to global AdS${}_4$ in vacuum Einstein-Hilbert gravity with negative cosmological constant.  These are non-compactly-generated black holes in which a single connected bulk horizon extends to meet the conformal boundary.  Thus the induced (conformal) boundary metric has smooth horizons as well.  In our examples, the boundary spacetime contains a pair of black holes connected through the bulk by a tubular bulk horizon.  Taking one boundary black hole to be hotter than the other ($\Delta T \neq 0$) prohibits equilibrium.    The result is a so-called flowing funnel, a stationary bulk black hole with a non-Killing horizon that may be said to transport heat toward the cooler boundary black hole. While generators of the bulk future horizon evolve toward zero expansion in the far future, they begin at finite affine parameter with infinite expansion on a singular past horizon characterized by power-law divergences with universal exponents.  We explore both  the horizon generators and the boundary stress tensor in detail.  While most of our results are numerical,
 a semi-analytic fluid/gravity description can be obtained by passing to a one-parameter generalization of the above boundary conditions.  The new parameter detunes the temperatures $T_\mathrm{bulk \ BH}$ and $T_\mathrm{bndy \ BH}$ of the bulk and boundary black holes, and we may then take $\alpha = \frac{T_\mathrm{bndy \ BH}}{T_\mathrm{bulk \ BH}}$  and $\Delta T$ small to control the accuracy of the fluid-gravity approximation.  In the small $\alpha, \Delta T$ regime we find excellent agreement with our numerical solutions. For our cases the agreement also remains quite good even for $\alpha \sim 0.8$.  In terms of a dual CFT, our $\alpha =1$ solutions describe heat transport via a large $N$ version of Hawking radiation through a deconfined plasma that couples efficiently to both boundary black holes.
\end{abstract}

 \maketitle

\newpage

\tableofcontents



\section{Introduction}
\label{intro}

We focus here on the classic problem of heat transport far from equilibrium, and away from the perturbative regime.  If the system of interest is an appropriate strongly coupled large $N$ conformal field theory (CFT), we may use gauge/gravity duality to exploit a perhaps-more-tractable description as a semi-classical bulk gravitational system.  We will consider the classical limit in cases where the bulk description may be truncated to $\Lambda < 0$ Einstein-Hilbert gravity. Our work complements perturbative computations of heat transport in this regime  (e.g. \cite{Son:2006em}), as well as non-perturbative studies of thermalization (see e.g. \cite{Chesler:2010bi,Heller:2011ju,Bantilan:2012vu} for recent examples and further references) and holographic shockwaves \cite{Khlebnikov:2010yt,Khlebnikov:2011ka}.

Suppose in particular that we couple a CFT in $d$ spacetime dimensions to heat sources or sinks of finite size and at finite locations.  A convenient way to introduce such sources is to place the CFT on a background non-dynamical spacetime containing stationary black holes with surface gravity $\kappa$, which have temperatures $\kappa/2\pi$ due to the Hawking effect.  As we review in section \ref{sec:detune} below, this problem may also be generalized so that the field theory temperature at the black hole horizon differs from $\kappa/2\pi$.  But since no information can flow outward across the horizon, the choice of a black hole metric is nevertheless useful to decouple our CFT from the details of the heat sources and sinks.  The problem of heat transport then becomes one of computing the expectation value of the stress tensor in the given background with the stated boundary conditions.  Since the background spacetime is not dynamical, we can choose the metric at will.  In particular, we can include as many black holes as we like at locations of our choosing, and we are free to assign their surface gravities as desired.  Of course, since we consider CFTs, we may also conformally rescale the background metric to reinterpret our heat sources/sinks as being infinitely large and located at infinite distance; more will be said about this alternate interpretation in section \ref{sec:detune} below.

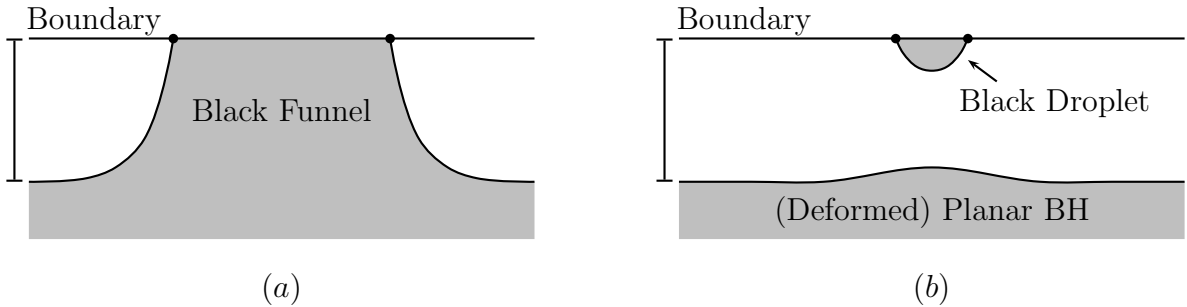
\begin{figure}[t]
\begin{center}
\hfill
\begin{pspicture}(-1,0)(16,5)
\psset{unit=0.95cm}


\pscustom{
\gsave
	\pscurve(0,2)(0.8,2.05)(1.2,2.2)(1.6,2.6)(2,4)
	\psline(5,4)
	\pscurve(5.4,2.6)(5.8,2.2)(6.2,2.05)(7,2)
	\psline(7,1.2)(0,1.2)(0,2)
	\fill[fillstyle=solid,fillcolor=lightgray]
\grestore
}

\psline(0,4)(7,4)
\psdots(2,4)(5,4)
\pscurve(0,2)(0.8,2.05)(1.2,2.2)(1.6,2.6)(2,4)
\pscurve(5,4)(5.4,2.6)(5.8,2.2)(6.2,2.05)(7,2)

\psline{|-|}(-0.2,2)(-0.2,4)

\rput[b](0.9,4.05){Boundary}
\rput(3.5,3){Black Funnel}
\rput(3.5,0.5){$(a)$}


\pscustom{
\gsave
	\pscurve(12,4)(12.1,3.8)(12.3,3.6)(12.5,3.55)(12.7,3.6)(12.9,3.8)(13,4)
	\psline(12,4)
	\fill[fillstyle=solid,fillcolor=lightgray]
\grestore
}

\pscustom{
\gsave
	\pscurve(9,2)(10,2)(11,2)(12.5,2.2)(14,2)(15,2)(16,2)
	\psline(16,1.2)(9,1.2)
	\fill[fillstyle=solid,fillcolor=lightgray]
\grestore
}

\psline(9,4)(16,4)
\psdots(12,4)(13,4)
\pscurve(12,4)(12.1,3.8)(12.3,3.6)(12.5,3.55)(12.7,3.6)(12.9,3.8)(13,4)
\pscurve(9,2)(10,2)(11,2)(12.5,2.2)(14,2)(15,2)(16,2)

\psline{|-|}(8.8,2)(8.8,4)

\psline{->}(13.4,3.4)(13,3.7)
\rput[t](14.2,3.3){Black Droplet}
\rput[b](9.9,4.05){Boundary}
\rput[t](12.5,1.8){(Deformed) Planar BH}
\rput(12.5,0.5){$(b)$}

\end{pspicture}
\caption{A sketch of the relevant solutions: {\bf (a)} a black funnel and  {\bf (b)} a black droplet above a deformed planar black hole. For simplicity, we take both  solutions to asymptote in the horizontal direction to the so-called planar AdS-Schwarzschild black hole.  As a result, both describe possible states of a CFT on an asymptotically flat black hole spacetime filled with a deconfined plasma at constant temperature.   In each figure, the top line corresponds to the spacetime on which the CFT lives; i.e., to the
conformal boundary of the AlAdS bulk.  The dots denote horizons of the boundary black holes.  The shading marks regions inside the bulk horizons. } \label{f:fundrop}
\end{center}
\end{figure}

Gauge/gravity duality for large $N$ field theories \cite{Maldacena:1997re} has been used to study related settings
in \cite{AM,Tanaka:2002rb,Emparan:2002px,Wiseman:2001xt,Wiseman:2002zc,Casadio:2002uv,Karasik:2003tx,Kudoh:2003xz,
Kudoh:2003vg,Kudoh:2004kf,Karasik:2004wk,Fitzpatrick:2006cd,Yoshino:2008rx,Gregory:2008br,
Hubeny:2009ru,Hubeny:2009kz,Hubeny:2009rc,Caldarelli:2011wa,Kleihaus:2011yq,Headrick:2009pv,
Figueras:2011va,Figueras:2011gd,Fischetti:2012ps,Santos:2012he}.  In this context, the $d$-dimensional black hole spacetime on which the CFT lives becomes the conformal boundary of a $(D=d+1)$-dimensional asymptotically locally anti-de Sitter (AlAdS) spacetime and we henceforth refer to our heat sources and sinks as boundary black holes.   Though the above explorations in gauge/gravity duality involved certain tensions and subtleties, the picture that emerged in \cite{Hubeny:2009ru} (building on \cite{Fitzpatrick:2006cd}) is one with two important phases for each boundary black hole, even when the CFT state is assumed to contain a deconfined plasma. See \cite{Fischetti:2012ps} for a condensed review.  In the so-called ``funnel phase'' a given boundary black hole is connected to distant regions of the boundary by a bulk horizon along which heat may be said to flow (say, if unequal temperatures are fixed at the two ends).  But there is no such connection in the contrasting ``droplet phase.''  Figure \ref{f:fundrop} depicts both phases for a simple case in which the boundary spacetime is asymptotically flat.  In the CFT description, the funnel phase allows a given boundary black hole to
exchange heat with distant regions much as in a free theory with a similar number of degrees of freedom.  One may say that grey body factors are ${\cal O}(1)$ even at large $N$.  But in the droplet phase there is no conduction of heat between a given black hole and the region far away at leading order in large $N$.   In effect, all grey body factors associated with the black hole vanish at this order\footnote{To be more precise, the grey body factors are non-zero only for a number of degrees of freedom that scales like $N^0=1$.}, so that the black hole does not couple efficiently to the surrounding plasma.  Additional phases are also possible that conduct heat between subsets of nearby black holes but not to infinity.

Until recently, both funnel and droplet solutions were largely conjectural. Explicit examples were known only in rather contrived settings or in low enough dimensions that all properties were determined by conformal invariance.  But numerical methods were used to construct more natural droplets in \cite{Figueras:2011va} and more natural funnels in \cite{Santos:2012he}.  An interesting detail is that the droplet solutions of \cite{Figueras:2011va} contain deformed planar black holes (see figure \ref{f:fundrop}) with vanishing temperature.  Constructing black droplet solutions that include finite-temperature (deformed) planar black holes remains an open technical challenge, though perturbative arguments give strong indications that they exist.

The above funnel and droplet papers largely focussed on cases without heat flow; i.e., either on droplets (in which heat does not flow in the approximation that the bulk is classical) or on equilibrium funnels.  The one exception was \cite{Fischetti:2012ps} which showed that, by changing conformal frames, rotating BTZ black holes \cite{Banados:1992wn,Banados:1992gq} in AdS${}_3$ can be re-interpreted as describing heat transport in 1+1 CFTs.  Here the standard left- and right-moving temperatures $T_L,T_R$  of the BTZ black hole correspond directly to the temperatures of the left- and right-moving components of the CFT.  Due to the strong constraints of conformal symmetry in low dimensions these components do not interact and the temperatures $T_L,T_R$ must be constants if the heat flux is stationary.  In addition, the flow of heat is necessarily isentropic (having no local generation of entropy).

We refer to black funnels transporting heat as ``flowing funnels."
Since none of the above special properties should hold for $d >2$ ($D > 3$), higher dimensional flowing funnels should be quite different than those found in \cite{Fischetti:2012ps}.  For example, a bulk horizon connecting two boundary black holes of different temperatures should (at least in some rough sense) be describable as having a temperature that varies along the horizon.  But recall that there is no generally accepted definition of horizon temperature which allows this temperature to vary\footnote{Except of course within the domain of the gradient expansion, as in the fluid-gravity correspondence \cite{Bhattacharyya:2008jc};  see also  \cite{Kinoshita:2011qs}.  For proposals in more general contexts see e.g. \cite{Abreu:2010ru} for a recent paper and references.  Our solutions may therefore provide interesting testbeds for such proposals.}.   Indeed, the fact that any definition of temperature should vary implies that the horizon is not Killing, which is already a novel property for a stationary black hole\footnote{For compactly generated horizons, this behavior is forbidden by the rigidity theorems \cite{R1,R2,Hollands:2006rj}.  But our bulk horizon is non-compact since it extends to the conformal boundary.}.  This suggests that the horizon generators have positive expansion (though of course tending to zero in the far future), so that they caustic at finite affine parameter in the past.   It is natural to expect this caustic to occur at a singular past horizon \cite{Hubeny:2009ru}, and section \ref{sec:results} confirms this picture for our solutions.

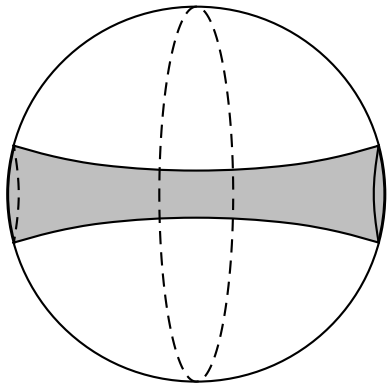
\begin{figure}[t]
\begin{center}
\psset{unit=1.25cm}
\begin{pspicture}(-3,-2)(3,2)

\pscircle(0,0){2}

\pscustom[fillstyle=solid,fillcolor=lightgray]{
\psarc(0,0){1.98}{-15}{15}
\pscurve[liftpen=1](1.932,0.518)(1.5,0.4)(1,0.31)(0,0.25)(-1,0.31)(-1.5,0.4)(-1.932,0.518)
\psarc(0,0){1.98}{165}{195}
\pscurve[liftpen=1](-1.932,-0.518)(-1.5,-0.4)(-1,-0.31)(0,-0.25)(1,-0.31)(1.5,-0.4)(1.932,-0.518)
}

\psellipticarc[linestyle=dashed](0,0)(0.4,2){7}{-7}
\psbezier(1.92,0.518)(1.85,0.1)(1.85,-0.1)(1.92,-0.518)
\psbezier[linestyle=dashed](-1.92,0.518)(-1.85,0.1)(-1.85,-0.1)(-1.92,-0.518)

\end{pspicture}
\caption{A sketch of a~$t = const.$ slice of the BTZ string~\eqref{eq:BTZstring}.  The two pieces of the boundary at~$z/\ell_3 = \pm \pi/2$ are conformal to two copies of the BTZ black hole, sketched above as the two hemispheres of an~$S^2$.  These boundary black holes are joined at infinity (the dashed line around the equator of the sphere), so that the boundary of the BTZ string can be thought of as a sphere with a black hole at either pole.  The bulk of the string is the interior of the sphere, where the string stretches from one black hole to the other.}
\label{fig:BTZstring}
\end{center}
\end{figure}

We focus below on what we call $D=4$ global flowing funnels, by which we mean deformations of the global AdS${}_4$ black string (also known as the Ba\~nados-Teitelboim-Zanelli (BTZ) black string; see e.g. \cite{Emparan:1999fd} where the solution was obtained as a special case of the AdS C-metric).  This reference solution may be constructed by starting with global AdS${}_4$ written in coordinates for which slices of constant radial coordinate $z$ are just AdS${}_3$.  One then replaces each such AdS${}_3$ slice with a BTZ metric \cite{Banados:1992wn,Banados:1992gq} having the correct $z$-dependence and which we chose to be nonrotating.  The result is an AlAdS Einstein metric which may be written
\begin{equation}
\label{eq:BTZstring}
ds^2 = \frac{\ell_4^2}{H^2(z)} \left[ -f(r) \, dt^2 + \frac{dr^2}{f(r)} + r^2 d\phi^2 + dz^2\right],
\end{equation}
with~$H(z) = \ell_3 \cos (z/\ell_3)$ and~$f(r) = (r^2 - r_0^2)/\ell_3^2$.  The solution is sketched in figure~\ref{fig:BTZstring}.  Here the horizon of the BTZ string is at~$r = r_0$, the parameter $\ell_4$ is the AdS$_4$ length scale, and the AdS$_3$ length scale~$\ell_3$ of the BTZ foliations may be set to any desired value by rescaling $z,r,r_0$.  The two boundary black holes (at $z/\ell_3 = \pm \pi/2$) have the same temperature, but we will seek deformations where these temperatures differ and heat flows between the two boundary black holes.

The outline of the paper is as follows.  Section \ref{sec:detune} reviews how static black funnels (i.e., without flow) may be generalized by adding a parameter $\alpha = T_{\rm bndy \ BH}/T_{\rm bulk \ BH}$ which allows the temperature of bulk and boundary horizons to differ.  In the small $\alpha$ limit, the analogous flowing funnels can be described in a derivative expansion; i.e., using the fluid/gravity correspondence of \cite{Bhattacharyya:2008jc}.  This correspondence is briefly reviewed and then applied to flowing funnels in section \ref{sec:fluid}.  Section \ref{sec:methods} then explains how to formulate the construction of flowing funnels with any $\alpha$ in a manner where one can proceed numerically.  The results of such numerics are presented in section \ref{sec:results} where they are compared to the fluid approximation of section \ref{sec:fluid}.  As expected, we find excellent agreement for small $\alpha$, though for our cases the agreement remains good even for $\alpha$ close to $1$.  We close with some final discussion in section \ref{sec:disc}.

Note: In the final stages of this work we learned of \cite{FW}, which also addresses the construction of AdS black holes with non-Killing horizons and may have some overlap with our work.  Their paper will appear simultaneously with ours on the arXiv.

\section{Detuning the bulk and boundary black hole temperatures}
\label{sec:detune}

As mentioned in the introduction, even without heat flow the black funnel paradigm may be generalized by adding an extra parameter $\alpha = \frac{T_{\rm bndy \ BH}}{T_{\rm bulk \ BH}}$ which allows us to detune to the temperatures of the bulk and boundary black holes.  In terms of the dual field theory, taking $\alpha \neq 1$ means that one considers a thermal ensemble at some temperature $T_{\rm Field \ Theory}$ which differs from the natural temperature $T_{\rm bndy \ BH}$ of the (say, static) boundary black hole spacetime on which the field theory lives.  One may think of the resulting state as defined by a Euclidean path integral with period $1/T_{\rm Field \ Theory} \neq 1/T_{\rm bndy \ BH}$ and thus having a conical singularity at the horizon of the boundary black hole.  What is interesting about this construction is that the gravitational dual can have a completely smooth Euclidean AlAdS bulk, with the conical singularity of the boundary geometry resulting only from a failure of the standard AlAdS boundary conditions at the singular boundary points \cite{Headrick:2010zt,Marolf:2010tg,Hung:2011nu}.  Any smooth horizon then clearly has temperature $T_{\rm bulk \ BH} = T_{\rm Field \ Theory} \neq T_{\rm bndy\ BH}$ as determined by the Euclidean period.

The prototypical detuned solution studied in \cite{Headrick:2010zt,Marolf:2010tg,Hung:2011nu} is just the general hyperbolic (sometimes referred to as `topological') black hole of \cite{Emparan:1998he,Birmingham:1998nr,Emparan:1999gf}, whose metric in $D=d+1$ bulk spacetime dimensions may be written
\begin{equation}
\label{eq:hyp}
ds^2_{d+1} = -F(r) dt^2 + \frac{dr^2}{F(r)} + r^2 d\Sigma^2_{d-1}, \ \ \ F(r) = \frac{r^2}{\ell^2_{d+1}} -1 - \frac{r_0^{d-2}}{r^{d-2}} \left(\frac{r^2_0}{\ell^2_{d+1}} -1 \right).
\end{equation}
Here $\ell_{d+1}$ is the AdS length scale associated with the ($D = d+1$)-dimensional cosmological constant, $d\Sigma^2_{d-1} =  d\xi^2 + \sinh^2 \xi d \Omega^2_{d-2}$ is the metric on the unit Euclidean hyperboloid, and $r=r_0$ is a smooth horizon of temperature
\begin{equation}
\label{eq:Htemp}
T_{\rm bulk \ BH} = \frac{r_0^2 \ell_{d+1}^{-2} d - (d-2)}{4\pi r_0}.
\end{equation}

Note that $F(r)$ approaches $r^2/\ell^2_{d+1}$ at large $r$.  Making an obvious choice of boundary conformal frame, the boundary metric is just the hyperbolic cylinder ${\bf H} \times {\mathbb R}$ with $ds^2_{{\bf H} \times {\mathbb R}} = -dt^2 + \ell^2_{d+1} d\Sigma_{d-1}$.  But note that we may write
\begin{equation}
\label{eq:hypbndy}
ds^2_{{\bf H} \times {\mathbb R}} =  - dt^2 + \frac{d\rho^2}{(1 - \rho^2/\ell_{d+1}^2)^2} + \frac{\rho^2}{1 - \rho^2/\ell_{d+1}^2} \, d\Omega_{d-2}^2,
\end{equation}
where $\rho/\ell_{d+1} = \tanh \xi$. Multiplying the right-hand side by $(1 - \rho^2/\ell_{d+1}^2)$ gives a metric on the static patch of the $d$-dimensional de Sitter space $dS_d$ with Hubble constant $\ell_{d+1}^{-1}$.  So by changing conformal frames in this way we may regard the boundary of \eqref{eq:hyp} as having de Sitter horizons with temperature $1/2\pi \ell_{d+1}$.  From the perspective of an observer in the static patch, the de Sitter horizon acts just like a black hole horizon with
\begin{equation}
\label{eq:Tbndy}
T_{\rm bndy \ BH} = \frac{1}{2\pi \ell_{d+1}}.
\end{equation}
For general $r_0$ this temperature clearly differs from that of the bulk horizon.  For the case where they agree, the hyperbolic black hole metric \eqref{eq:hyp} is just pure AdS${}_{d+1}$ in appropriate hyperbolic coordinates.  We recall that even for the tuned case $\alpha =1$ ref. \cite{Santos:2012he} found the conformal frame \eqref{eq:hypbndy} useful for constructing black funnel solutions numerically.

Since the analysis of temperatures above involves only the horizons, it is clear that detuned bulk and boundary horizons should exist much more generally.
Indeed, any static, spherically symmetric boundary metric with a pair of of smooth horizons at $\rho = \pm \ell_{d+1}$ may be written in the form
\begin{equation}
\label{eq:GenbndyBH}
ds^2_{\rm bndy \ BH} =  \left(1 - \rho^2/\ell_{d+1}^2\right) \left( - \widetilde{F}(\rho) dt^2 + \frac{d\rho^2}{\widetilde{G}(\rho)} + \widetilde{R}^2(\rho) d\Omega_{d-2}^2 \right),
\end{equation}
where $\widetilde F$,~$\widetilde{G}$, and~$1/\widetilde R^2$ are smooth on some interval including $\rho \in [-\ell_{d+1}, \ell_{d+1}]$,  $\widetilde G$ has a second order zero at each of $\rho = \pm \ell_{d+1}$, and $1/\tilde R^2$ vanishes at $\rho = \pm \ell_{d+1}$.  So after a conformal transformation \eqref{eq:GenbndyBH} agrees with \eqref{eq:hypbndy} to leading order in $\rho$ for each term and in this sense may be said to approach ${\bf H} \times {\mathbb R}$ at large $\rho$.  The ansatz \eqref{eq:GenbndyBH} can equivalently be written
\begin{equation}
\label{eq:GenbndyBH2}
ds^2_{\rm bndy \ BH} =  e^{-2x/x_0} F(x)  \left( -dt^2 + dx^2  + R^2(x) d\Omega_{d-2}^2 \right),
\end{equation}
where~$x_0$ is some reference length scale and $F$ and $e^{\mp 2x/x_0}R^2$ are smooth functions of $e^{\mp 2x/x_0}$ at  $e^{\mp 2 x/x_0} = 0$.  In particular, up to the conformal factor $ds^2_{{\bf H} \times {\mathbb R}}$ takes this form for $x_0 = \ell_{d+1}$ and $R^2 = \ell_{d+1}^2 \sinh^2(x/\ell_{d+1})$.  In terms of \eqref{eq:GenbndyBH2} the boundary black holes have temperatures
\begin{equation}
\label{eq:Tbndy2}
T^\mathrm{bndy \ BH}_\pm = \frac{1}{4\pi} \, \lim_{x \rightarrow \pm \infty} \frac{d}{dx} \ln \left(e^{-2x/x_0}F(x)\right).
\end{equation}

It is therefore sensible to choose any $r_0$ and seek a smooth bulk solution in which each term approaches that of \eqref{eq:hyp} to leading order in $e^{-2|x|/x_0}$ at large $|x|$; see section \ref{sec:methods} for a more complete analysis of these boundary conditions.  Any static such solution will have a bulk horizon with temperature \eqref{eq:Htemp} and can again be interpreted as being dual to a field theory state of this temperature on a black hole background of temperature \eqref{eq:Tbndy}.  In the next sections we will seek further generalizations with different values of $r_0$ (which we then call $r_\pm$) at $x = \pm \infty$.  That is to say that for $ x \rightarrow + \infty$ the bulk solution will asymptote as above to \eqref{eq:hyp} with $r_0 = r_+$, while for
$ x \rightarrow - \infty$ it will analogously approach \eqref{eq:hyp} with $r_0 = r_-$.  The bulk horizon may then be said to approach the temperatures $T_\pm$ given by \eqref{eq:Htemp} with $r_0$ replaced $r_\pm$.  We will also allow distinct temperatures $T_\pm^{\rm bndy \ BH}$ for the $x = \pm\infty$ boundary black holes and introduce the parameters $\alpha_\pm = T_\pm^{\rm bndy
\ BH}/T_\pm$.  In fact, we will always take $\alpha_+ = \alpha_-$.

Of course, we may also consider so-called ultrastatic conformal frames analogous to  \eqref{eq:hypbndy}.    Starting with~\eqref{eq:GenbndyBH2} and multiplying by a conformal factor~$e^{2x/x_0}/F(x)$, one obtains the boundary metric
\begin{equation}
\label{eq:ultrastatic}
ds^2 =  -dt^2 + dx^2  + R^2(x) d\Omega_{d-2}
\end{equation}
for which $\partial_t$ is a hypersurface-orthogonal Killing field of norm $-1$.
In this frame, the boundary spacetime has two asymptotic regions, each asymptotic to ${\bf H} \times {\mathbb R}$ (say, with the same curvature scale $\ell_{d+1}$).  Furthermore, in the CFT description each region contains an infinite reservoir of deconfined plasma.  Such infinite reservoirs may act as heat baths, and indeed the boundary conditions imply that they are in thermal equilibrium at temperature $T_\pm$ in the limits $x \rightarrow \pm \infty$.

\section{The fluid limit}
\label{sec:fluid}

While a general treatment of black funnels remains challenging, it is by now well known that the study of AdS black holes simplifies in the so-called hydrodynamic limit of the fluid/gravity correspondence \cite{Bhattacharyya:2008jc} in which all other parameters vary slowly in comparison with the black hole temperature and the solution can be described using a derivative expansion.  For any fixed boundary metric, taking the limit of large temperature (i.e., small $\alpha_\pm$) makes all metric derivatives small in this sense. We may thus expect a good hydrodynamic description if in addition we control temperature gradients by taking $\Delta T = T_+ - T_-$ small.

The key point in the analysis of \cite{Bhattacharyya:2008jc} is that, having chosen a boundary conformal frame with boundary metric $g^{(0)}_{ij}$, every AlAdS${}_{d+1}$ solution is associated with a $d$-dimensional boundary stress tensor $T_{ij}$ which is traceless and conserved on the boundary:
\begin{equation}
\label{bst}
 g^{(0)}_{ij} T^{ij} =0,   \ \ \ D_i T^{ij} =0,
\end{equation}
where $D_i$ is the covariant derivative compatible with $g^{(0)}_{ij}$ .  Below, we use the boundary metric $g^{(0)}_{ij}$ and its inverse to raise and lower indices $i,j,k,l,\dots$.

As an example, consider the planar AdS-Schwarzschild black hole
\begin{equation}
ds^2_\mathrm{AdS-Schw} = -\left(r^2/\ell_{d+1}^2\right) \left(1-r_0^d/r^d\right) \, dt^2 + \frac{\ell_{d+1}^2 dr^2}{r^2 \left(1-r_0^d/r^d\right)} + \left(r^2/\ell_{d+1}^2\right) d\mathbf{x}_{d-1}^2,
\end{equation}
with $r_0 = 4\pi \ell_{d+1}^2 T/d$.  Taking the boundary metric to be $ds^2_\mathrm{bndy} = -dt^2 + d\mathbf{x}_{d-1}^2$, one finds
\be
\label{eq:Tideal}
T^{ij} = T^{ij}_\mathrm{ideal} = \rho u^i u^j + P \, \mathcal{P}^{ij},
\ee
which takes the form of an ideal fluid with velocity field $u^i\partial_i = \partial_t$, transverse projector $\mathcal{P}^{ij} = g^{ij} + u^i u^j$, and
\be
\label{eq:EOS}
\rho = (d-1)\frac{\T^d}{16\pi \ell_{d+1} G^{(d+1)}} ,\ \ \ P =  \frac{\rho}{d-1},
\ee
which of course satisfies \eqref{bst}.
In \eqref{eq:EOS}, we have defined for convenience~$\T \equiv 4\pi \ell_{d+1} T/d$.  By a simple Lorentz transformation we may obtain corresponding solutions with any constant (normalized) timelike $d$-velocity $u^i$.

The main result of \cite{Bhattacharyya:2008jc} was to show that the temperature $T$ and $d$-velocity $u^i$ may be promoted to slowly-varying functions of the boundary coordinates $\mathbf{x}, t$ (at which point we refer to them collectively as the hydrodynamic fields).  Here the term ``slowly-varying'' is defined with respect to the temperature as measured in the local rest frame selected by $u^i$.  In particular, under these conditions \cite{Bhattacharyya:2008jc} showed that a smooth bulk solution may be constructed via a gradient expansion so long as $u^i$ is everywhere timelike and the associated boundary stress tensor does indeed satisfy \eqref{bst}.  They further showed that at each order in this expansion the conditions \eqref{bst} may be expressed as standard hydrodynamic equations for a (conformal) fluid with velocity field $u^i$, which we take to satisfy $u^i u_i =-1$.  This last step essentially just repeats the standard derivation of hydrodynamics from conservation laws.

In particular, ref. \cite{Bhattacharyya:2008jc}  showed that the boundary stress tensor takes the form
\be
\label{eq:gradient}
T^{ij} = T^{ij}_\mathrm{ideal} + \sum_{n=1} \Pi_{(n)}^{ij},
\ee
where~$\Pi_{(n)}$ are dissipative terms that are~$n^\mathrm{th}$ order in derivatives of the hydrodynamic fields; for example,
\be
\label{eq:P1}
\Pi_{(1)}^{ij} = -2\eta\sigma^{ij},
\ee
where
\be
\label{eq:shearvisc}
\eta = \frac{\T^{d-1}}{16\pi G^{(d+1)}}
\ee
is the shear viscosity, and~$\theta = D_i u^i$ and
\be
\label{eq:shear}
\sigma^{ij} = \mathcal{P}^{ik} \mathcal{P}^{jl} D_{(k} u_{l)} - \frac{\theta}{d-1} \,  \mathcal{P}^{ij}
\ee
are respectively the divergence and shear of the velocity field.  In writing \eqref{eq:P1} there is a freedom to make certain field redefinitions which, following \cite{Bhattacharyya:2008jc},  we have removed by choosing the so-called Landau frame in which the $\Pi_{(n)}^{ij}$ are taken to be purely transverse.

Since by assumption derivatives of the hydrodynamic fields are parametrically small in some parameter~$\eps$,~$\Pi_{(n)}$ is of order~$\eps^n$.  Below, we solve the fluid equations \eqref{bst} at order $n=0$ and $n=1$ for the ultrastatic boundary metrics \eqref{eq:ultrastatic} and a purely radial velocity field (so that the only non-vanishing components are $u^t$,~$u^x$).  We also assume the flow to be stationary, so that $u^i$,~$\T$ are independent of time.

A new effect at first order is the appearance of dissipation, and thus the production of entropy.  At zeroth order, the entropy current $J_S^i$ takes the simple form~$(J_S^i)_\mathrm{ideal} = s u^i$, where~$s(x) = \T^{d-1}/4G^{(d+1)}$ is the entropy density.  Using the equations of motion and thermodynamic relations, one can show~\cite{Rangamani:2009xk} that
\begin{equation}
D_i \left(J_S^i\right)_\mathrm{ideal} = 0.
\end{equation}
At first order, the entropy current still takes the form~$(J_S^i)_1 = s u^i$, but its divergence now becomes~\cite{Rangamani:2009xk}
\begin{equation}
\label{eq:sgen}
D_i \left(J_S^i\right)_1 = \frac{8\pi \ell_{d+1}\eta}{d\T} \, \sigma_{ij} \sigma^{ij} \geq 0,
\end{equation}
showing that entropy is produced unless $\sigma_{ij}=0$.

\subsection{Ideal Fluid}
\label{sec:zero}

We begin at order $n=0$.  We denote the associated fluid quantities $\T_0, u_0^i$ and work in~$d=3$.  Following~\cite{Das:2010mk}, we project the fluid equations into components parallel and perpendicular to the velocity.   These yield respectively
\begin{equation}
\label{eq:01}
D_i \left(\T^2_0 u^i_0\right) = 0 \ \ \ {\rm and} \ \ \  D^k \T_0 + u^i_0 D_i\left(\T_0 u^k_0\right) = 0,
\end{equation}
or
\begin{equation}
\label{eq:02}
\partial_x \left(\sqrt{-g^{(0)}} \, \T^2_0 u^x_0\right) = 0   \ \ \ {\rm and} \ \ \  \partial_x \left(\T_0 \left(u_0\right)_t\right) = 0.
 \end{equation}
 Thus
\begin{equation}
\T^2_0 u^x_0 = \frac{\T_\infty^2}{2aR},  	 \ \ \ \T_0 \left(u_0\right)_t = \T_\infty,
\end{equation}
in terms of integration constants that we have chosen to call
$\T_\infty^2/2a$, $\T_\infty$.  Since $u^2 =-1$, it remains to solve a quadratic equation for $\T_0, u^i_0$.  We of course obtain two solutions labeled by a choice of sign.  The solution with finite and nonzero asymptotic temperatures $T_\pm$ has
\begin{align}
\label{eq:T0}
\T^2_0 &= \frac{\T_\infty^2}{2}\left[1 + \sqrt{1-\frac{1}{a^2 R^2}}\right], \\
\label{eq:u0}
u^x_0 &= aR\left[1 - \sqrt{1-\frac{1}{a^2 R^2}}\right].
\end{align}
Note that since $R$ diverges at large $x$, at this order the asymptotic temperatures $T_\pm$ at $x\rightarrow \pm \infty$ agree; i.e.,  $\Delta T = T_+ - T_- = {\cal O}(\epsilon)$.  We also find $u^x_0 \rightarrow 0$ at $x = \pm \infty.$

\subsection{First Order Corrections}
\label{sec:firstorder}

To compute corrections to~\eqref{eq:T0},~\eqref{eq:u0}, we choose to solve the fluid equations \eqref{bst} iteratively. Introducing a bookkeeping parameter~$\eps$ to keep track of derivatives, we may write $\T = \T_m + {\cal O}(\eps^{m+1})$,~$u^i = u^i_m + {\cal O}(\eps^{m+1})$ for each $m$.  We compute $\T_m$,~$u^i_m$ by dropping terms with $n > m$ in \eqref{eq:gradient} and evaluating the remaining $\Pi_{(n)}^{ij}$ on $\T_{m-n}$,~$u^i_{m-n}$.  Thus $\T_m$,~$u^i_m$ enter \eqref{bst} only through $T^{ij}_\mathrm{ideal}$ and the equations to be solved are essentially just \eqref{eq:01}, \eqref{eq:02} with additional source terms given by the $\Pi_{(n)}^{ij}$.   The integration constants (as well as the sign choices that come from solving quadratic equations) may be fixed by requiring $\T_m$,~$u_m^i$ to approximate $\T_{m-1}$,~$u_{m-1}^i$ to the desired order as $\eps \rightarrow 0$.

To first order, one finds
\begin{align}
\label{eq:T1}
\T_1^2 &= \frac{1}{2}\left(B(x) + \T_\infty\right)^2 \left[1 + \sqrt{1 - \left(\frac{2(A(x) + \T_\infty^2/2a)}{(B(x) + \T_\infty)^2 R}\right)^2}\right], \\
\label{eq:ux1}
u^x_1 &= \frac{(B(x) + \T_\infty)^2 R}{2(A(x) + \T_\infty^2/2a)} \left[1 - \sqrt{1 - \left(\frac{2(A(x) + \T_\infty^2/2a)}{(B(x) + \T_\infty)^2 R}\right)^2}\right],
\end{align}
where
\begin{align}
A(x) &= \frac{2\ell_4}{3} \int_0^x R \left[\T \sigma^{ij} \sigma_{ij} \right]^{(0)} \, dx', \\
B(x) &= \frac{2\ell_4}{3} \int_0^x \left[\frac{\T^{-2} D_i \left(\T^2 {\sigma^i}_t \right) - u_t \sigma^{ij} \sigma_{ij}}{u^x}\right]^{(0)} \, dx',
\end{align}
and the square brackets~$[\cdot]^{(0)}$ indicate that the enclosed quantities are evaluated on the zeroth order solutions~\eqref{eq:T0},~\eqref{eq:u0}.  At this order, the asymptotic temperatures differ and are given by the (finite) expression
\be
\T(\pm \infty) = \T_\infty + B(\pm \infty),
\ee
so that
\be
\Delta \T := \T(\infty) - \T(-\infty) = B(\infty) - B(-\infty).
\ee

It is useful to consider the further limit of small $\Delta \T$, which greatly simplifies the above results.  This is equivalent to taking $a$ large.  Since
\be
B(x) = \frac{2\ell_4}{3}\int_0^x \left[-\frac{R''(x')}{2R^2(x')} \, \frac{1}{a} + \mathcal{O}\left(\frac{1}{a}\right)^2 \right] \, dx',
\ee
we find
\be
\Delta \T = \frac{2\ell_4}{3}\int_{-\infty}^\infty \left[-\frac{R''(x')}{2R^2(x')} \, \frac{1}{a} + \mathcal{O}\left(\frac{1}{a}\right)^2 \right] \, dx' = -\frac{I}{3a} + \mathcal{O}\left(\frac{1}{a}\right)^2,
\ee
for
\be
I := \ell_4 \int_{-\infty}^\infty \frac{R''(x)}{R^2(x)} \, dx.
\ee
Noting that~$A(x) = \mathcal{O}(1/a)^2$ we then find
\begin{align}
u^t_1 &= 1 + \mathcal{O}(\Delta \T^2), \\
u^x_1 &= -\frac{3 \Delta \T}{2IR(x)} + \mathcal{O}(\Delta \T^2), \\
\T_1 &= \T_\infty + \frac{\ell_4 \Delta \T}{I} \, \int_0^x \frac{R''(x')}{R^2(x')} \, dx' + \mathcal{O}(\Delta \T^2),
\end{align}
so that the non-zero components of the stress tensor are
\begin{align}
16\pi \ell_4 G^{(4)} {T^t}_t &= -2\T_\infty^3 - \frac{6\ell_4 \T_\infty^2 \Delta \T}{I} \, \int_0^x \frac{R''(x')}{R^2(x')} \, dx' + \mathcal{O}(\Delta \T^2), \\
16\pi \ell_4 G^{(4)} {T^t}_x &= -\frac{9\T_\infty^3 \Delta \T}{2IR(x)} + \mathcal{O}(\Delta \T^2), \\
16\pi \ell_4 G^{(4)} {T^x}_x &= \T_\infty^3 - \frac{3\ell_4 \T_\infty^2 \Delta \T}{I}\, \left(\frac{R'(x)}{R^2(x)} - \int_0^x \frac{R''(x')}{R^2(x')} \, dx'\right) + \mathcal{O}(\Delta \T^2), \\
16\pi \ell_4 G^{(4)} {T^\phi}_\phi &= \T_\infty^3 + \frac{3\ell_4 \T_\infty^2 \Delta \T}{I}\, \left(\frac{R'(x)}{R^2(x)} + \int_0^x \frac{R''(x')}{R^2(x')} \, dx'\right) + \mathcal{O}(\Delta \T^2).
\end{align}

Note that the lowest order term in the energy flux~$T^{tx}$ is linear in~$\Delta\T$; this naturally leads to a notion of thermal conductivity.  We first calculate the heat flux~$\Phi$ as the energy flux integrated over a circle of constant~$x$:
\be
\label{eq:flux}
\Phi = 2\pi R(x)  T^{tx} = -\frac{9\T_\infty^3 \Delta \T}{16 \ell_4 G^{(4)} I} + \mathcal{O}(\Delta \T^2).
\ee
We define the thermal conductivity as~$k := -d\Phi/d\Delta T|_{\Delta T = 0}$ so that
\be
\label{eq:thermcond}
k = \frac{3\pi \T_\infty^3}{4 G^{(4)} I}.
\ee

We have also explored the analogous results at second order $n=2$ in the hydrodynamic approximation.  While the general expressions are unenlightening,  each quantity above agrees with the $n=1$ expression up to linear order in~$\Delta\T$ for all $\T_\infty$.  In particular, the conductivity $k$ is unchanged.

Finally, the entropy current~$\left(J_S^i\right)_1 = su^i$ for our solutions is
\begin{align}
4G^{(4)} \left(J_S^t\right)_1 =& \ \frac{(B(x) + \T_\infty)^4 R}{2\sqrt{2}\,(A(x) + \T_\infty^2/2a)}\left[1-\sqrt{1 - \left(\frac{2(A(x) + \T_\infty^2/2a)}{(B(x) + \T_\infty)^2 R}\right)^2}\right]^{1/2} \\ &\times \left(1+\sqrt{1 - \left(\frac{2(A(x) + \T_\infty^2/2a)}{(B(x) + \T_\infty)^2 R}\right)^2}\right), \\
4G^{(4)} \left(J_S^x\right)_1 =& \ \frac{1}{R}\left(A(x) + \frac{\T_\infty^2}{2a}\right),
\end{align}
which has divergence
\be
4G^{(4)} D_i \left(J_S^i\right)_1 = \frac{2\ell_4}{3} \left[\T \sigma_{ij} \sigma^{ij}\right]^{(0)} = \frac{\ell_4\T_\infty}{3\sqrt{2}} \frac{R'^2}{R^2(a^2R^2 - 1)}\left[1 + \sqrt{1-\frac{1}{a^2 R^2}}\right]^{1/2}.
\ee
To lowest nonvanishing order in~$\Delta\T$, these become
\begin{align}
4G^{(4)} \left(J_S^t\right)_1 &= \T_\infty^2 + \frac{2\ell_4 \T_\infty\Delta \T}{I} \, \int_0^x \frac{R''(x')}{R^2(x')} \, dx' + \mathcal{O}(\Delta \T^2), \\
4G^{(4)} \left(J_S^x\right)_1 &= -\frac{3 \T_\infty^2 \Delta\T}{2IR(x)} + \mathcal{O}(\Delta \T^2),
\end{align}
\be
\label{eq:eglin}
4G^{(4)} D_i \left(J_S^i\right)_1 = \frac{3\ell_4\T_\infty \Delta \T^2}{I^2} \, \frac{(R')^2}{R^4} + \mathcal{O}(\Delta\T)^3.
\ee
Note that the divergence of the current is of order~$\Delta\T^2$ as expected from \eqref{eq:sgen}.
It turns out that \eqref{eq:eglin} is unchanged when one passes to second order in the hydrodynamic expansion, though the entropy current $J_S^i$ itself changes even at zeroth order in $\Delta \T$.

These expressions may of course be transformed to any other conformal frame.  The ultrastatic frame~\eqref{eq:ultrastatic} used above had the convenient feature that, at least at small velocity, the local fluid temperature  (defined with respect to proper time in the fluid rest frame)  coincides with the temperature defined with respect to the static Killing field~$\partial_t$.  In a more general conformal frame, these two temperatures do not coincide even at small velocity.  Note that we will employ only time-independent conformal transformations below, so that $\partial_t$ remains a Killing field in all frames.  We will continue to refer to temperatures normalized (up to a boost to the fluid rest frame) with respect to $\partial_t$  by~$\T$, while we denote the local fluid temperature (defined with respect to rest-frame proper time) as~$\T_\mathrm{loc}$.   Thus $\T$ is unchanged by the conformal transformation while $\T_\mathrm{loc}$ is rescaled.

For comparison with our later numerics, appendix \ref{app:bhf} presents the results in the black hole frame for the explicit metric functions and in terms of the particular coordinates used in section \ref{sec:results} below.  The resulting more explicit expressions are correspondingly more complicated than those above.

\section{How to flow a more general funnel}
\label{sec:methods}

Our family of flowing funnels will be labeled by four parameters: the temperatures $T^\mathrm{bndy \ BH}_\pm$ of the left- and right- boundary black holes and the temperatures $T_{\pm}$ associated with the left- and right- ends of the bulk black hole.  As discussed in section \ref{sec:detune} these four temperatures are completely independent in principle, though in our simulations we will always set $\alpha_+ = \alpha_-$ which introduces one relation.

The most generic ansatz compatible with our symmetry requirements depends on seven unknown functions:
\begin{multline}
\label{eq:ansatz}
ds^2 = \frac{\ell_4^2}{(1-w^2)^2(1-y^2)^2}\left\{-M(y)G(w)^2(1-w^2)^2\,y^2\,A\,\left[\ell_4^{-1}d\tilde{t}+Q(w)\frac{\chi_2}{y}dy\right]^2\right. \\
\left.+\frac{4(1-w^2)^2\,B\,dy^2}{M(y)}+y^2_0\left[\frac{4\,S_1}{2-w^2}\,\left(dw+\ell_4^{-1} \chi_1\,d\tilde{t}+\frac{F\,dy}{y}\right)^2+S_2 d\phi^2\right] \right\}\,,
\end{multline}
where $A$, $B$, $F$, $S_1$, $S_2$, $\chi_1$ and $\chi_2$ are all functions of $y$ and $w$. In addition we have defined
\be
G(w) = 1+\frac{\beta}{2} w^3(5-3\,w^2)\,,\quad M(y) = 2-y^2-\frac{(1-y^2)^2(1-y^2_0)}{y^2_0}\quad\text{and}\quad Q(w)=1+\frac{2}{M(0)G(w)}.
\label{eq:aux}
\ee
The insertion of these factors will be justified later, when we will also see that $\beta$ controls the temperature difference between the two boundary black holes, and $y_0$ is a parameter that controls the validity of the fluid approximation. Here $y$ ranges over $[0,1]$ and $w$ ranges over $[-1,1]$, with $y=0$ being the bulk horizon and $y=1$ the conformal boundary. At least for $y \neq 0$ regions with $w \sim \pm 1$ are close (in the sense of a conformal diagram) to where either bulk horizon meets either the left or right boundary black hole (compare with figure \ref{fig:BTZstring}). As we will explain below, the symbol $\tilde t$ was used in \eqref{eq:ansatz} in order to save $t$ for another coordinate associated with Fefferman-Graham gauge.  However, $\partial_{\tilde t} = \partial_t$ so we will refer to the time-translation as simply $\partial_t$.

\subsection{Boundary Conditions}

At the conformal boundary ($y=1$) we impose the boundary conditions
\be
\label{eq:bcyh}
A(w,1)=B(w,1)=S_1(w,1)=S_2(w,1)=\chi_2(w,1)=1,\quad F(w,1)=\chi_1(w,1) =0,
\ee
which ensures a boundary metric conformal to
\be
\label{eq:BBH}
\ell_4^{-2} ds^2_\partial = - \frac{1}{\ell_4^2 y^2_0}(1-\hat \rho^2)^2 G(\hat \rho)^2\,dt^2+\frac{4 d\hat \rho^2}{2-\hat \rho^2}+d\phi^2\, ,
\ee
where $\hat \rho = \rho/\ell_4$.
As in section \ref{sec:detune}, we refer to \eqref{eq:BBH} as the boundary metric in the black hole conformal frame. In presenting our results in section \ref{sec:results} we will describe all boundary quantities, such as the stress energy tensor, with respect to this frame. The boundary metric $ds^2_\partial$ has horizons at $\hat \rho=\pm1$ with Hawking temperatures
\be
T^\mathrm{bndy \ BH}_\pm = \frac{G(\pm1)}{2\pi \ell_4 y_0}\,.
\ee
 We will extract the boundary stress tensor following the strategy of \cite{Santos:2012he} and using the results of \cite{deHaro:2000xn}. The only technical difference with respect to \cite{Santos:2012he} involves the relation between the coordinates $(\tilde{t},w,y,\phi)$ and Fefferman-Graham coordinates $(t,z,\rho,\phi)$. Due to the cross term $\chi_2$ in Eq.~(\ref{eq:ansatz}) the map between $\tilde{t}$ and $t$ is not trivial, instead it is expressed as a powers series in $z$ of the form:
\be
\tilde{t}=t+z\,T_1(\rho)+\mathcal{O}(z^2)
\ee
where for instance $T_1(\rho) = -Q(\rho)y_0/(2(1-\hat{\rho}^2))$.

The left and right boundaries lie at $w=\pm1$.  There we impose
\begin{equation}
\label{eq:bchyper}
A(\pm1,y)=B(\pm1,y)=S_1(\pm1,y)=S_2(\pm1,y)=\chi_2(\pm1,y)=1,\quad F(\pm1,y)=\chi_1(\pm1,y) =0,
\end{equation}
which reduces Eq.~(\ref{eq:ansatz}) to
\begin{multline}
\label{eq:linehyperbc}
ds^2|_{w\to\pm1} = \frac{\ell_4^2}{(1-y^2)^2}\Bigg\{-M(y)G(\pm1)^2\,y^2\left[\ell_4^{-1} d\tilde{t}+Q(\pm1)\frac{dy}{y}\right]^2+\frac{4\,dy^2}{M(y)}\\+\frac{y^2_0}{(1\mp w)^2}\left(dw^2+\frac{d\phi^2}{4}\right) \Bigg\}\,.
\end{multline}
Under the coordinate transformation:
\begin{equation}
y=\sqrt{1-\frac{r_0}{r}},\quad \frac{r_0}{\ell^2_4}d\tau=G(\pm1)\left( \ell_4^{-1}d\tilde{t}+\frac{Q(\pm1)\,dy}{y}\right)\,,\quad w = \pm 1\mp e^{-\xi}\,,\quad y_0\equiv \frac{r_0}{\ell_4},
\end{equation}
the line element (\ref{eq:linehyperbc}) yields the large $\xi$ limit of Eq.~(\ref{eq:hyp}) with $d=3$. The fact that our ansatz (\ref{eq:ansatz}) reduces to a hyperbolic black hole at $w=\pm1$ displays the physical meaning of $y_0$ as an overall scale that controls the bulk horizon temperatures (and thus $\alpha_\pm$). Note that the line element (\ref{eq:linehyperbc}) also defines $T_{\pm} = T^\mathrm{bndy \ BH}_\pm M(0)/2$. If $y_0 = 1$, then $T_{\pm} = T^\mathrm{bndy \ BH}_\pm$, \emph{i.e.} it represents the `tuned' case $\alpha_\pm = 1$. Thus the fluid approximation becomes more accurate as $y_0$ increases, or equivalently, as $\alpha_\pm$ decrease.

We have imposed Dirichlet data at each of the above three edges of our computational domain. But it remains to specify boundary conditions at $y=0$, the flowing funnel horizon. Here we demand that the line element (\ref{eq:ansatz}) be smooth in ingoing Eddington-Finkelstein coordinates (which cover the future horizon). To understand the explicit form of this condition, we introduce local ingoing Eddington-Finkelstein coordinates ($v,\tilde w, \tilde y, \phi$) through
\be
dv = d\tilde{t}+\ell_4 \frac{d\tilde{y}}{2 \tilde{y}}+\mathcal{O}(\tilde{y}^0),\quad d\tilde{w}=\frac{d w}{\chi_1(w,0)}+\ell_4^{-1} dv+\mathcal{O}(\tilde{y}^0)\,,\quad y=\tilde{y}^{1/2}.
\label{eq:coordinatestilde}
\ee
The terms omitted in the above $\tilde{y}$ expansion can be chosen such that a line of constant $(v, \tilde w, \phi)$ is an ingoing null geodesic. Note that lines of constant $v$ have $d\tilde{y}/d\tilde{t}<0$, as required for ingoing coordinates. Furthermore, regularity of the line element (\ref{eq:ansatz}) in the above coordinates requires
\begin{multline}
F(w,0)=\chi_1(w,0)\,,\quad B(w,0)=\frac{M(0)^2G(w)^2A(w,0)}{4}\left[1-Q(w)\chi_2(w,0)\right]^2 \\
\partial_y A(w,0)=0\,,\quad \partial_y S_1(w,0)=0\,,\quad \partial_y S_2(w,0)=0\,,\quad \partial_y \chi_1(w,0)=0\,,\quad \partial_y \chi_2(w,0)=0\,.
\label{eq:boundaryconditionshorizon}
\end{multline}
We will find $\chi_1(w,0)$ to be finite and non-zero (at $w \neq \pm 1$), so our original $w$ is already an ingoing coordinate.  It will thus be straightforward to read off results associated with the future horizon.

The past horizon is more subtle.   It is located at $v\to-\infty$ and can be reached along lines of constant $\tilde{w}$.  Depending on the sign of $\chi_1$, this tends to drive $w$ to either $\pm 1$.  Below, we consider $T_+ > T_-$ so that the hotter black hole is on the right.  One might therefore expect $w$ to decrease along the horizon generators so their coordinate velocity is toward the cooler black hole; i.e., one might expect $\chi_1(w,0) > 0$.  But for the particular ansatz we have chosen our numerics turn out to give $\chi_1 < 0$ (see section \ref{sec:results}) so that the past horizon in fact lies at $w=-1$.  This appears to be a coordinate artifact, though a full understanding is beyond the scope of this work.

Below, we will solve the Einstein equations (with cosmological constant) in the form
\begin{equation}
E_{ab} := R_{ab}+\frac{3}{\ell^2_4}g_{ab}=0,
\label{eq:einsteinR}
\end{equation}
subject to the boundary conditions detailed above.

\subsection{The DeTurck Method}

The diffeomorphism invariance of \eqref{eq:einsteinR} means that these equations do not lead to a well-posed boundary value problem.  While one could attempt to proceed by gauge-fixing,  a clever trick known as the DeTurck method was introduced in \cite{Headrick:2009pv} and in \cite{Figueras:2011va,Adam:2011dn} was shown to succeed (under rather general assumptions) when one seeks appropriate stationary equilibrium solutions of the vacuum Einstein equations,  with or without a negative cosmological constant. Though our situation turns out to fall outside the bounds of the proof given in \cite{Figueras:2011va}, we nevertheless employ this method successfully below.

We begin with a brief review.
The DeTurck method is based on the so called Einstein-DeTurck equation
\begin{equation}
E^{H}_{ab}\equiv E_{ab}-\nabla_{(a}\hat{\xi}_{b)}=0,
\label{eq:turck}
\end{equation}
which differs from from Eq.~(\ref{eq:einsteinR}) by the addition of $-\nabla_{(a}\hat{\xi}_{b)}$.
Here  $\hat{\xi}^a=g^{cd}[\Gamma^a_{cd}(g)-\Gamma^a_{cd}(\bar{g})]$,  $\Gamma(\mathfrak{g})$ is the Levi-Civita connection associated with the metric $\mathfrak{g}$, and $\bar g$ is some specified non-dynamical reference metric. Since $\hat{\xi}$ is defined by a difference between two connections, it transforms as a tensor.  Hereafter $\bar{g}$ will be chosen to have the same asymptotics and horizon structure as $g$. In particular, it must satisfy the same Dirichlet boundary conditions as $g$.

Clearly any solution to $E^H_{ab} = 0$ with $\hat{\xi}= 0$ also solves $E_{ab} = 0$. But one may ask if \eqref{eq:turck} can have additional solutions that do not satisfy $E_{ab} = 0$. Under a variety of circumstances one can show that solutions with $\hat{\xi} \neq 0$, the so called Ricci solitons, cannot exist \cite{Figueras:2011va}. However, the assumptions used in \cite{Figueras:2011va} seem not to hold for our system of equations. In particular, after reduction along the symmetry directions $t,\phi$ our system turns out to have a mixed-elliptic hyperbolic nature. This is most easily seen from the fact that, while our system will be elliptic near infinity where $\partial_t$ is timelike,  we expect an ergoregion near the horizon where all linear combinations of $\partial_t,\partial_\phi$ are spacelike. So in this region reduction along $(t,\phi)$ gives a Lorentz-signature metric on the base space.  This differs qualitatively from the case of Kerr, where $\partial_t, \partial_\phi$ span a timelike plane everywhere outside the horizon and reduction along $(t,\phi)$ gives a Euclidean-signature metric on the base space. See \cite{Adam:2011dn} for a more detailed discussion. The difference arises  from the fact that the Kerr horizon `flows' only along the Killing field $\partial_\phi$ while our horizon `flows' in the $w$ direction, which is not associated with any symmetry.
Thus Ricci solitons may well exist in our case.  But for any solution to (\ref{eq:turck}) one may simply calculate $\hat{\xi}$ to see if it vanishes.  For all of our flowing funnel solutions discussed below we find $\hat{\xi}=0$ to machine precision.

It remains to specify our choice of reference metric $\bar{g}$.  We choose $\bar{g}$ to be given by the line element (\ref{eq:ansatz}) with $A=B=S_1=S_2=\chi_2 =1$ and $F=\chi_1 =0$. This enforces all Dirichlet boundary conditions except those at the horizon, Eq.~(\ref{eq:boundaryconditionshorizon}). To satisfy these remaining conditions we need only choose $Q(x)$ as in Eq.~(\ref{eq:aux}).

\subsection{Numerical Method}

\noindent We use a standard pseudospectral collocation approximation in $w$, $y$ and solve the resulting non-linear algebraic equations using a damped Newton method with damping monitoring function
$|\hat{\xi}_t|$.  This ensures that Newton's method takes a path in the approximate solution space that decreases $|\hat{\xi}_t|$ at each step.  This method may also prove useful in solving more general mixed elliptic-hyperbolic systems.  We represent the $w$ and $y$ dependence of all functions as a series in Chebyshev polynomials. As explained above, our integration domain lives on a rectangular grid, $(w,y)\in[-1,1]\times[0,1]$.

\begin{figure}[t]
\centerline{
\includegraphics[width=0.4\textwidth]{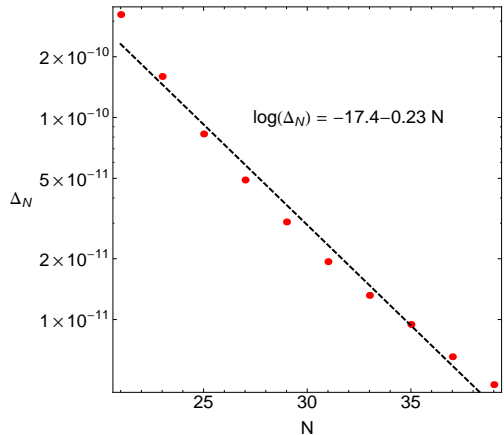}
}
\caption{$\Delta_N$ as a function of the number of grid points $N$. The vertical scale is logarithmic, and the data is well fit by an exponential decay: $\log(\Delta_N) = -17.4-0.23\,N$.}
\label{fig:convergence}
\end{figure}

To monitor the convergence of our method we have computed the total heat flux $\Phi$ (defined by the first equality in \eqref{eq:flux}) for several resolutions. We denote the number of grid points in $w$ and $y$ by $N$ and compute $\Delta_N = |1-\Phi_N/\Phi_{N+1}|$ for several values of $N$. The results for this procedure are illustrated in Fig.~\ref{fig:convergence} for $\beta = 0.1$ and $y_0 = 1$. We find exponential convergence with $N$, as expected for pseudospectral collocation methods. Furthermore, in order to ensure that we are converging to an Einstein solution rather than a Ricci soliton we monitor all components of $\hat{\xi}$. For all plots shown in this manuscript, each component of $\hat{\xi}_a$ has absolute value smaller than $10^{-10}$.

\section{Results and comparisons}
\label{sec:results}

\begin{figure}[t]
\centering
\subfigure[]{
    \includegraphics[width=0.4\textwidth]{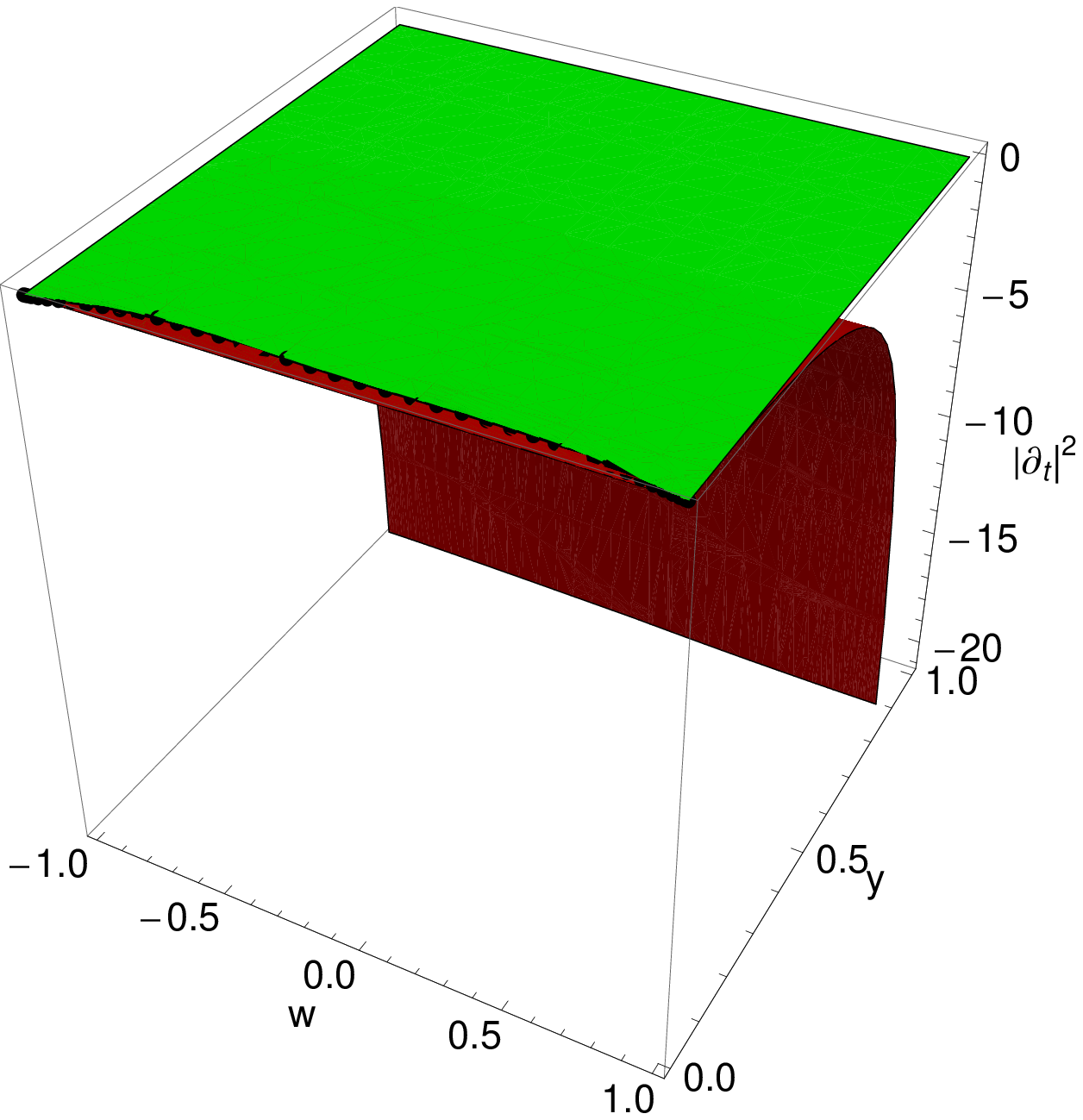}
    \label{fig:normdtspacetime}
}
\subfigure[]{
    \includegraphics[width=0.4\textwidth]{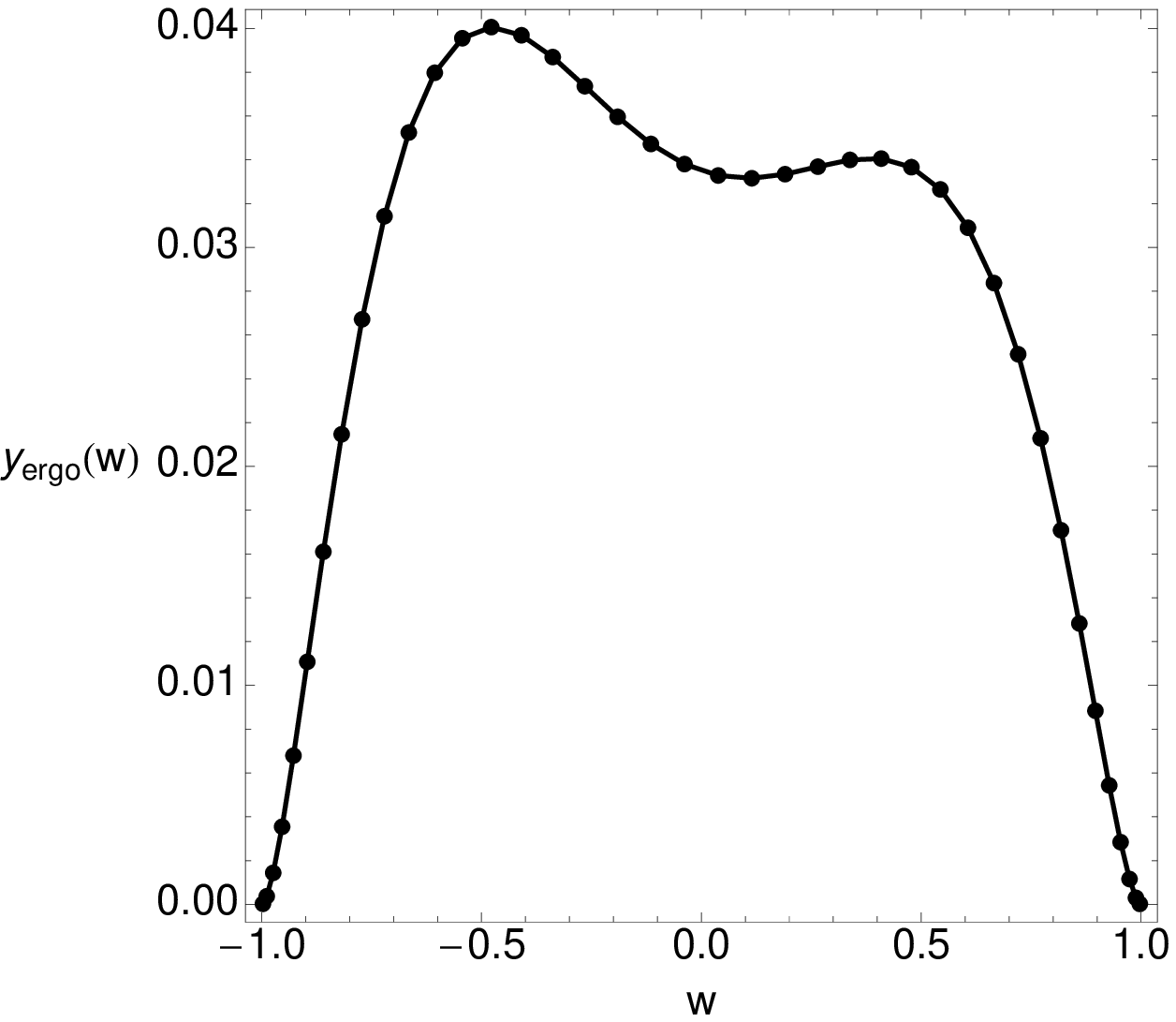}
    \label{fig:ergo}
}
\caption[Optional caption for list of figures]{(a): The curved surface shows the  norm of $\partial_t$ over our integration domain. To guide the eye, we also draw a flat horizontal surface at zero norm.  (b): The ergosurface as a function of $w$.  Both figures use $\alpha_\pm = 1$ and $\Delta T/T_{\infty} = 0.2$.}
\label{figs:normsdt1}
\end{figure}

We now present the results of our numerical analysis and compare them with the first-order ($n=1$) hydrodynamic approximation. The plots below are labeled by a parameter $T_\infty$ whose definition
\begin{equation}
T_\infty = \frac{\left[256 \left(144 \sqrt{2}-557\right) \lambda ^2+105 \pi  \left(293 \lambda ^2+128\right)\right] \left(3 y_0^2-1\right)}{28 \pi  \left[15 \pi  \left(293 \lambda^2+128\right)-11008 \lambda ^2\right] y^2_0}
\end{equation}
was inspired by the first-order hydrodynamic result \eqref{eq:T1}.  For small $\Delta T$ we have $T_\infty = (T_+ + T_-)/2 + {\cal O}(\Delta T)^2$.  We note that all the numerical results we will present use units where~$\ell_4 = 1$ (so that~$\rho = \hat{\rho}$) and~$16\pi G^{(4)} = 1$.  We also take $T_+ > T_-$ so that the hotter black hole lies on the right.

\begin{figure}[h]
\centering
\subfigure[]
{
 \includegraphics[width=0.42\textwidth]{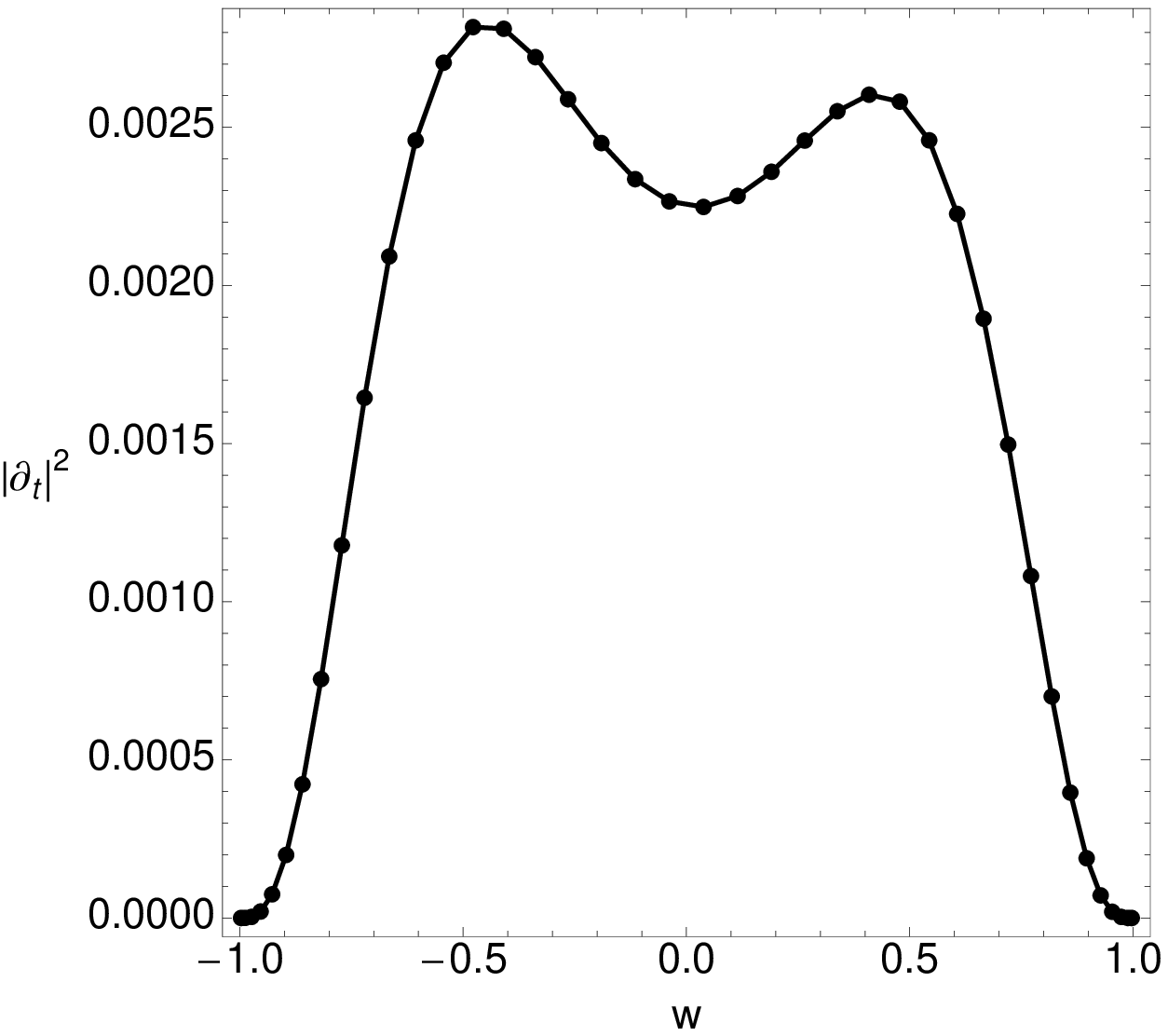}
}
\subfigure[]
{
 \includegraphics[width=0.4\textwidth]{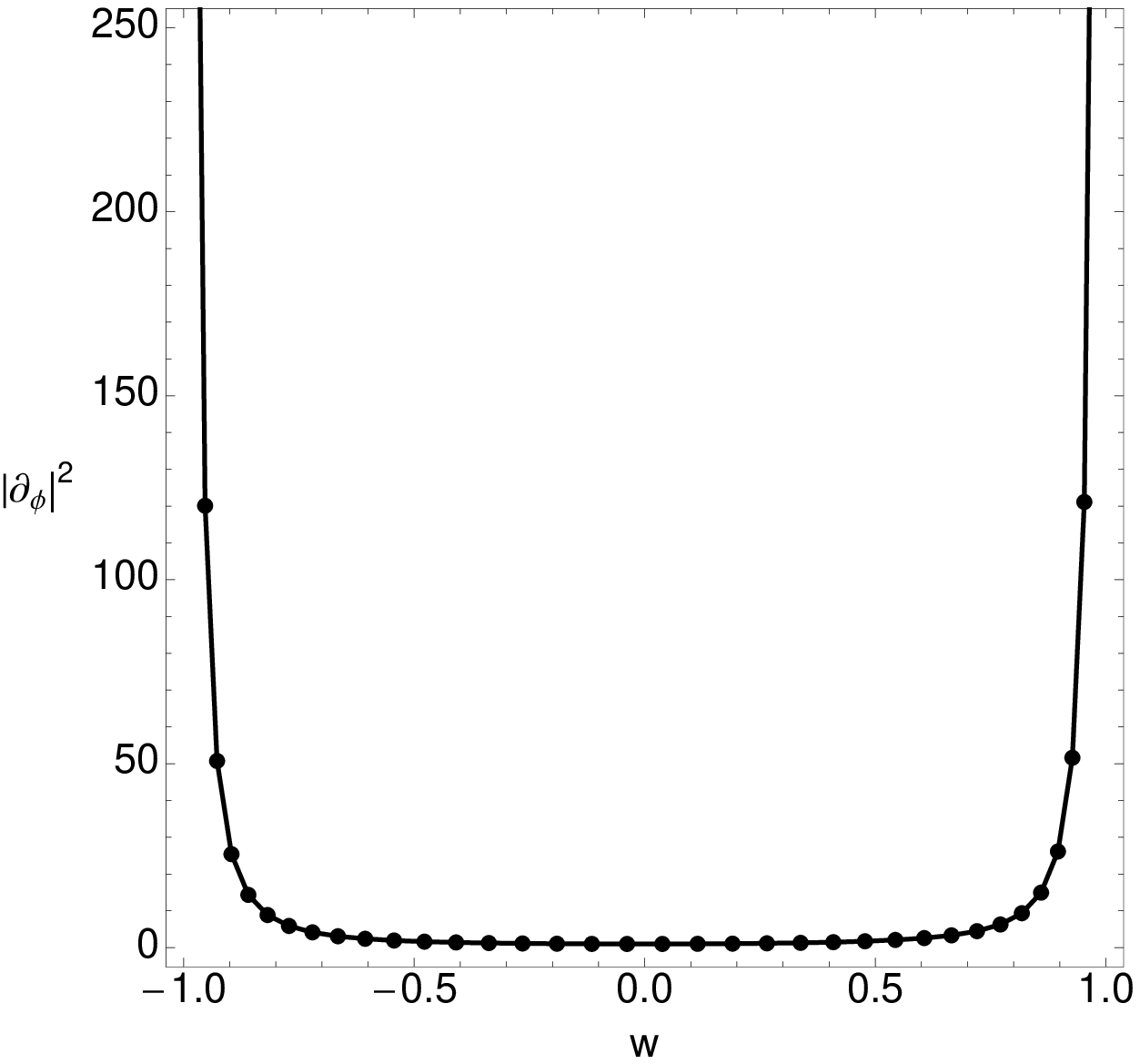}
 \label{fig:hcomponents}
}
\label{fig:normhorizon}
\caption[Optional caption for list of figures]{(a): The norm $|\partial_t|^2$ on the future horizon.  (b): The norm $|\partial_\phi|^2$ on the future horizon.  Both figures use $\alpha_\pm = 1$ and $\Delta T/T_{\infty} = 0.2$ and are plotted as functions of $w$. }
\end{figure}

We begin with the norm $|\partial_t|^2$ of the time translation.  Figure~\ref{fig:normdtspacetime} shows a typical plot. To guide the eye we have also plotted a reference surface of constant $|\partial_t|^2=0$.  The two surfaces intersect at the ergosurface, whose location we display separately in Fig.~\ref{fig:ergo}. Inside the ergoregion $|\partial_t|^2$ becomes positive, changing the character of Eq.~(\ref{eq:turck}) from elliptic to hyperbolic. This region is at the core of the difficulties in trying to prove that our numerical method ensures $\hat{\xi}=0$ on solutions of Eq.~(\ref{eq:turck}) with appropriate boundary conditions. Fig.~\ref{fig:normhorizon} shows $|\partial_t|^2$ and, for comparison and later use, $|\partial_\phi|^2$  as a function of $w$ along the horizon.  We remind the reader that $\partial_t$ and $\partial_\phi$ are precisely orthogonal everywhere in our spacetime, so this describes the full induced metric $h_{IJ}$ (for $I,J = t, \phi$) in the 2-plane spanned by $\partial_t, \partial_\phi$. Both norms are clearly positive everywhere on the horizon, though $|\partial_t|^2$ never becomes very large even with $\Delta T/T_{\infty} = 0.2$.   This may help to explain why our numerical approach succeeded.

\begin{figure}[t!]
\centerline{
\includegraphics[width=0.9\textwidth]{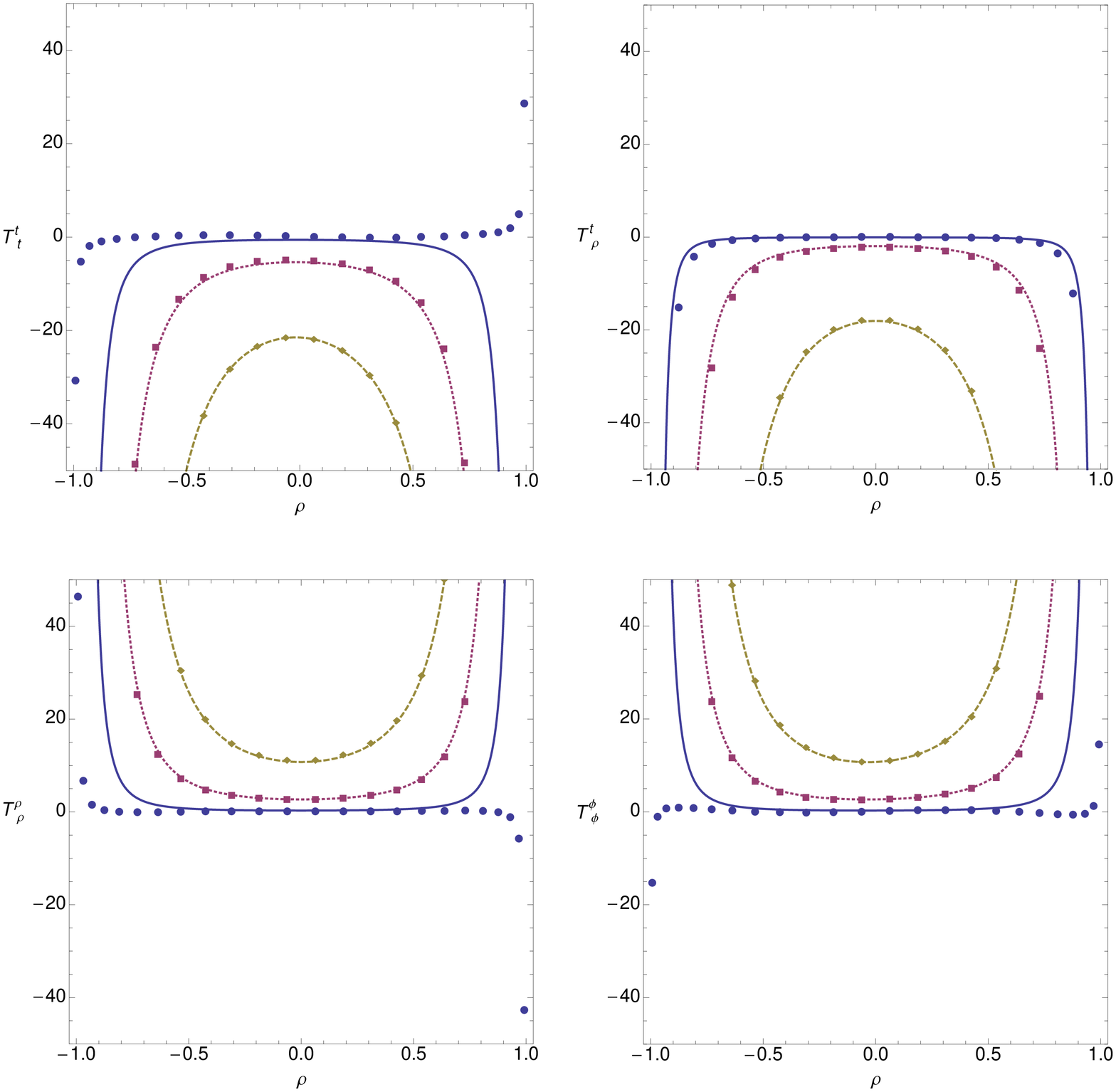}
}
\caption{Components of the boundary stress energy tensor as a function of $\rho$ for fixed $\beta = 0.04$. Each panel shows the first-order ($n=1$) hydrodynamic prediction as lines and the exact numerical data as symbols. The disks and solid line show $\alpha_\pm = 1$, the squares and dashed line show $\alpha_\pm = 0.77$ and the diamonds and dotted line show $\alpha_\pm = 0.70$. These corresponds to $\Delta T/T_\infty = 0.080$, $\Delta T/T_\infty = 0.050$ and $\Delta T/T_\infty = 0.034$, respectively.
Since $\Delta T/T_\infty$ is small, we have used only the linear results from appendix \ref{app:bhf} to plot the hydrodynamics.}
\label{fig:stressenergytensor}
\end{figure}

Let us now discuss the behavior of the boundary stress tensor. For small $\alpha, \Delta T/T_\infty$, this quantity may also be computed using the hydrodynamic approximation of section \ref{sec:fluid} and provides another good check of our numerics.    Fig.~\ref{fig:stressenergytensor} shows the components of the stress energy tensor as a function of the boundary coordinate $\rho$ for several values of $\alpha$ at fixed  $\beta$. The lines represent the first order hydrodynamic prediction and the symbols represent data extracted from our numerics. Large stress tensors correspond to larger values of $\alpha$.   We see that at least for small $\Delta T/T_{\infty}$ the fluid gravity prediction works remarkably well even for for $\alpha \sim 0.8$. The agreement of all of these curves when $\alpha$ is small is a reassuring test of our numerics.  However, at larger $\alpha$ qualitative differences from our hydrodynamic approximation begin to appear.  For example, we note that while $T^t{}_{t}$ is always negative (and thus the energy density is positive) in the hydrodynamic regime,  for $\alpha \gtrsim 1$ our simulations show $T^t{}_{t}$ becomes positive near the hotter black hole.

\begin{figure}[t!]
\centerline{
\includegraphics[width=0.5\textwidth]{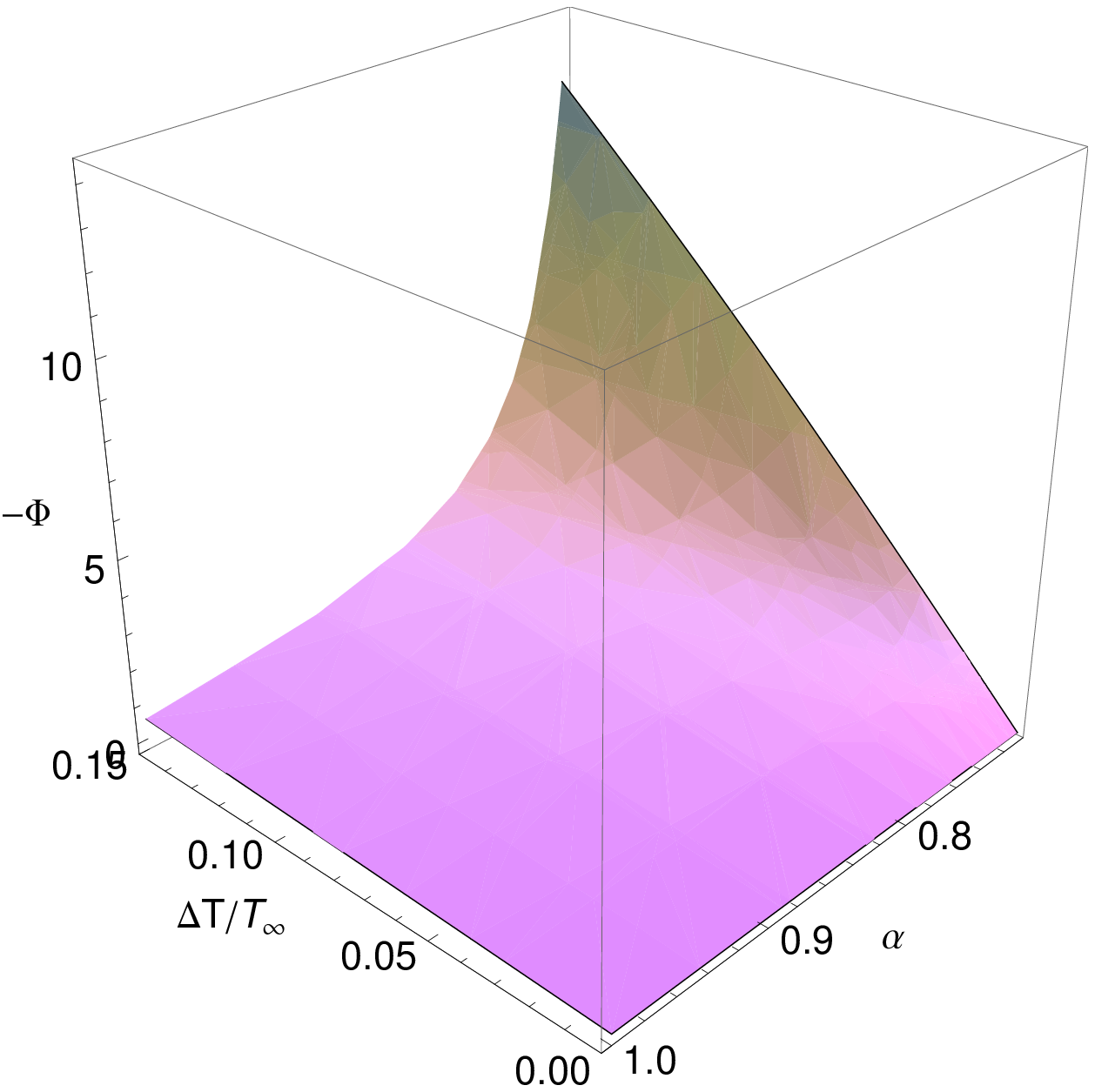}
}
\caption{Three-dimensional plot of the boundary flux extracted from our numerics as a function of $\Delta T/T_{\infty}$ and $\alpha = \alpha_+ = \alpha_-$.}
\label{fig:boundaryflux3d}
\end{figure}
\begin{figure}[h!]
\centerline{
\includegraphics[width=0.9\textwidth]{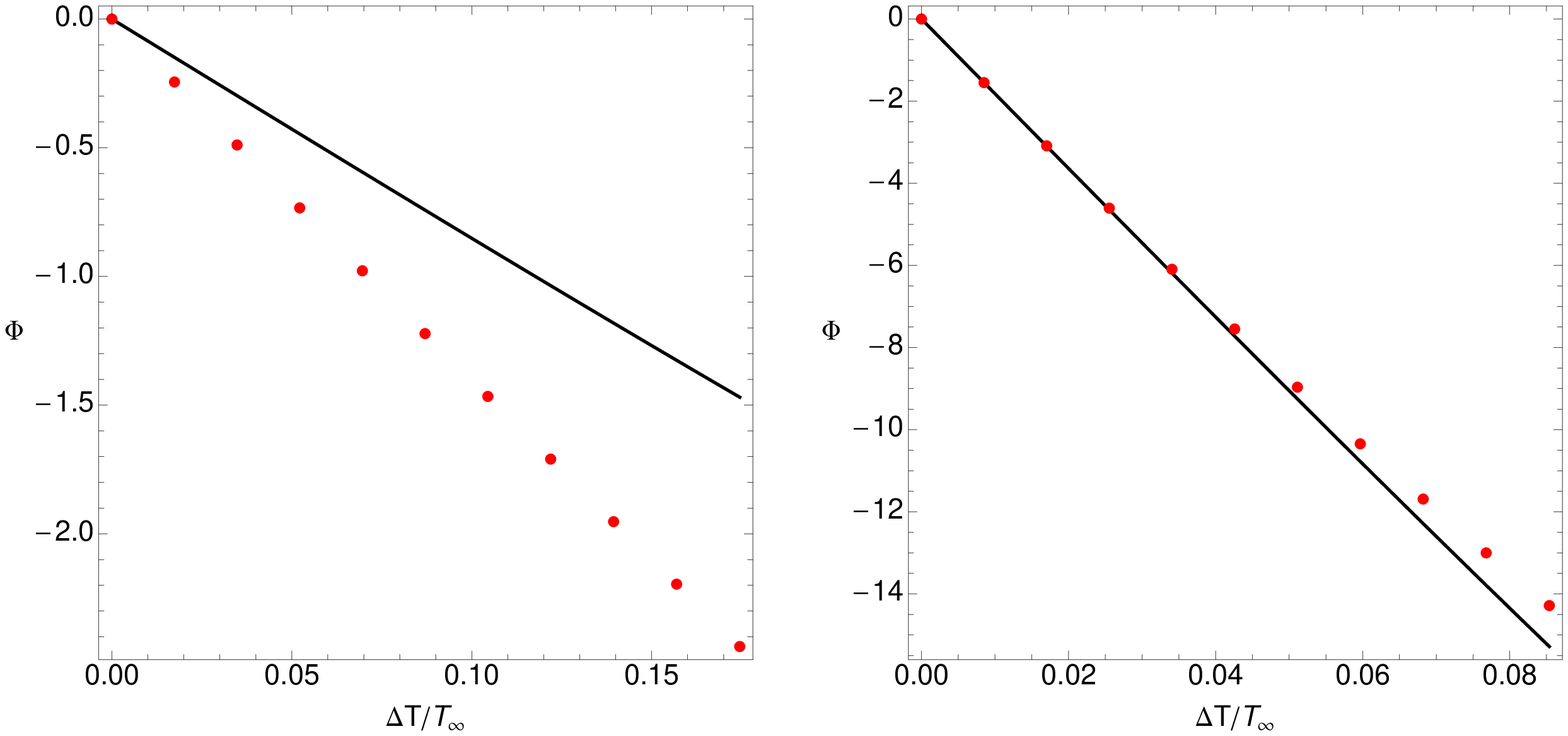}
}
\caption{The total heat flux $\Phi$ as a function of $\Delta T/T_{\infty}$ for $\alpha = 0.9$ (left) and $\alpha  = 0.7$ (right). The solid curves are the first order hydrodynamic results.  Since $\Delta T/T_\infty$ is small, we have used only the linear results from appendix \ref{app:bhf}. The dots show our numerical data. }
\label{fig:boundaryflux}
\end{figure}

From the standpoint of the dual CFT, the main physical result of our paper is displayed in Fig.~\ref{fig:boundaryflux3d}.  This plot shows how the total heat flux $\Phi$ varies for different values of $\Delta T/T_{\infty}$ and $\alpha = \alpha_+ = \alpha_-$. We see that it increases in magnitude as $\Delta T/T_{\infty}$ increases, and also as $\alpha$ decreases.  This computation can be seen as a first principle calculation for the thermal conductivity of a strongly coupled plasma at large $N$ beyond the hydrodynamic regime.  Fig.~\ref{fig:boundaryflux} compares some~$\alpha = \mathrm{const.}$ cross-sections of Fig.~\ref{fig:boundaryflux3d} to the the results of first-order ($n=1$) hydrodynamics at linear order in $\Delta T$; i.e., to \eqref{eq:linBHF1}-\eqref{eq:linBHF4}.  These show good agreement for small~$\alpha$ and $\Delta T$, but deviate as expected at larger $\alpha$.

It remains to examine the horizon more closely. Our horizon is a three-dimensional null surface and, since $\partial_t, \partial_\phi$ are both spacelike and tangent to the horizon, any two null geodesics that generate the future horizon generators are related by some isometry.  Thus all generators are equivalent, though it remains to understand the evolution of the spacetime along each generator. We compute the affine parameter, expansion, and shear along each generator using simple expressions in terms of the induced metric $h_{IJ}$ (for $I,J=t,\phi$) on the 2-plane spanned by $\partial_t,\partial_\phi$.  These expressions are given in appendix \ref{sec:horizon}. We study each of these quantities only on the surface $y=0$.

\begin{figure}[t]
\centering
\subfigure[]{
    \includegraphics[width=0.4\textwidth]{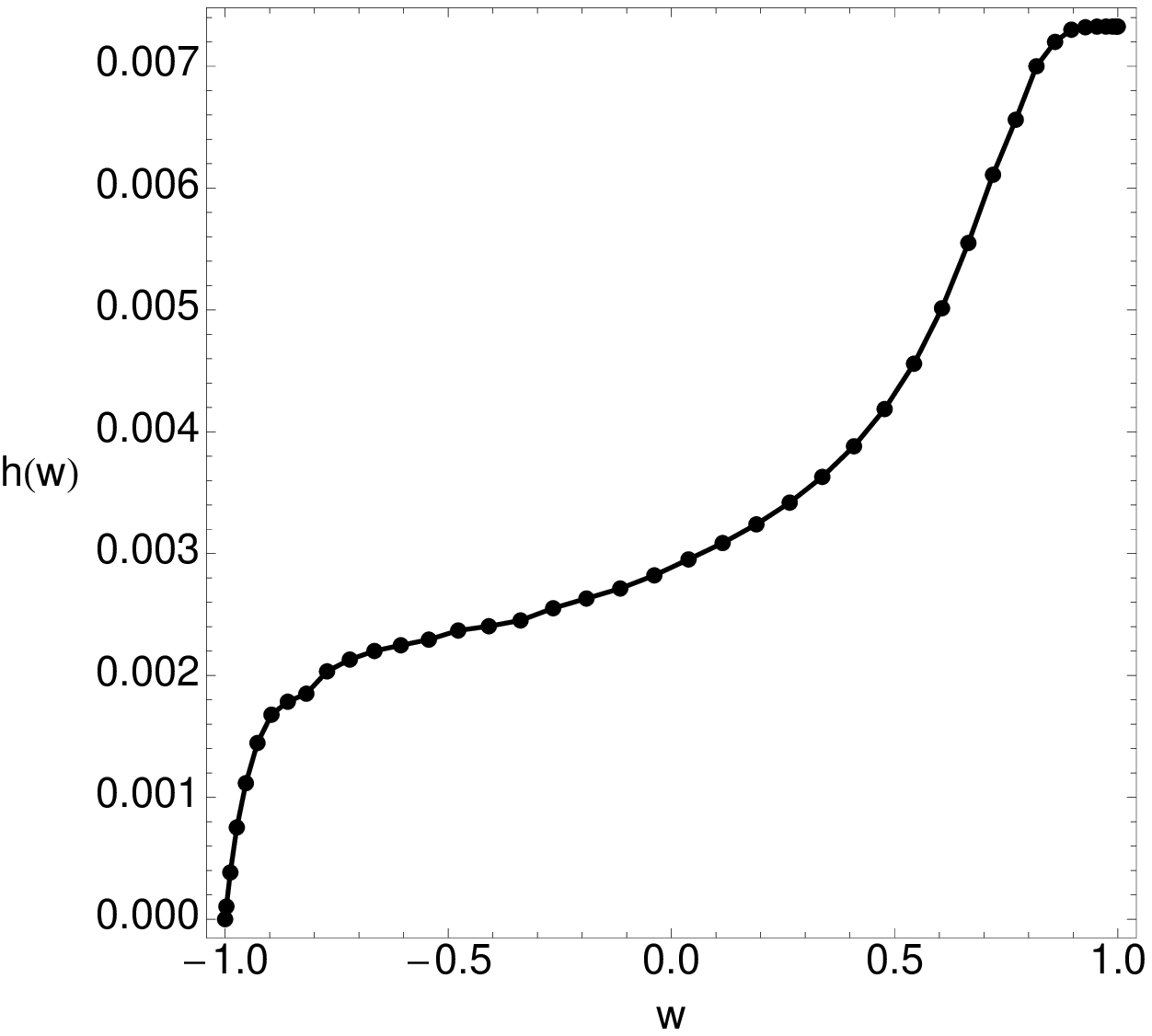}
\label{fig:hdet}
}
\subfigure[]{
    \includegraphics[width=0.42\textwidth]{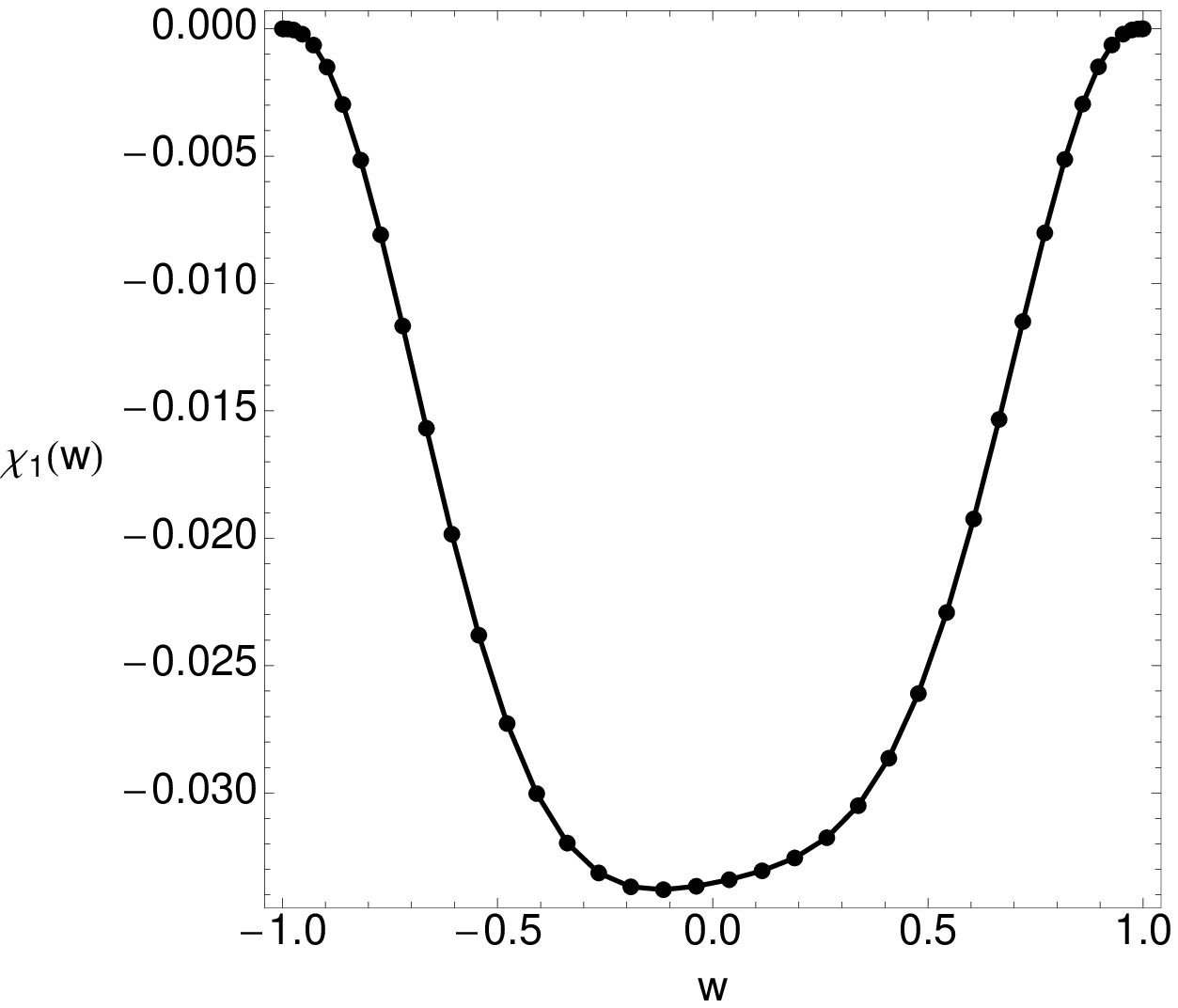}
    \label{fig:chi1}
}
\caption[Optional caption for list of figures]{(a): The determinant $h = h_{tt} h_{\phi \phi}$. (b): The metric component $\chi_1$ along the horizon.  Both quantities are plotted as functions of $w$ for $\alpha_\pm = 1$ and $\Delta T/T_{\infty} = 0.2$.}
\label{figs:normsdt2}
\end{figure}

We begin with $h_{IJ}$ itself. Recall that $w = \pm1$ are the asymptotic regions of static hyperbolic black holes where the Killing field $\partial_t$ becomes null at the horizon and $|\partial_\phi|^2$ becomes large.  These behaviors are clearly shown in figure \ref{fig:hdet}.  But these similarities between $w=\pm 1$  are misleading and the actual behaviors at $w = \pm 1$ are quite different. This may be seen from the plot of $h = \det{h_{IJ}} = h_{tt}h_{\phi \phi} = |\partial_t|^2 |\partial_\phi|^2$ in Fig.~\ref{fig:hdet}.  This determinant vanishes at $w=-1$ but approaches a non-zero constant at $w=+1$.  Note that $h$  is monotonic along $y=0$, as it must be along a smooth horizon.

What is perhaps surprising is that $h$ is an increasing function of $w$.  This shows that $w$ increases toward the future along the future horizon, so that the past horizon must lie at $w=-1$. In contrast, in the coordinates of e.g. \cite{Bhattacharyya:2008xc}, the coordinate velocity of the horizon generators would be in the direction of heat transport, and thus (since we take the cooler black hole to lie at $w=-1$) toward negative $w$.  Standard coordinates for Kerr also behave like those of \cite{Bhattacharyya:2008xc} and have the equivalent of our $\chi_1$ being positive for positive angular velocity.
In contrast, we find $\chi_1$ to be negative at the horizon; see figure \ref{fig:chi1}.  Since $\chi_1$ samples completely different metric components than $h$, we take this as a strong indication that our solutions are consistent despite the surprising location of the past horizon.  Another strong indication of consistency is the above agreement between our boundary stress tensors and those predicted by the hydrodynamic approximation.  Indeed, we have tested for various possible errors (such as inverting the sign of $\Delta T$) in our code by examining the effect of various sign changes on Fig.~\ref{fig:stressenergytensor} and found in each case that such changes would lead to notable discrepancies with hydrodynamics.  In particular, we stress that our simulations give the physically correct sign for the heat flux ${T^t}_\rho$.

The apparent proximity of the past horizon to the cooler black hole must thus be a coordinate artifact. We have confirmed this expectation by repeating our simulations in the coordinates defined by Eq.~(\ref{eq:coordinatestilde}) and finding that the equivalent of $\chi_1$ is positive for negative heat flux. 
For comparison, we mention that also note that a similarly surprising sign can be found in the 2+1 flowing funnels of \cite{Fischetti:2012ps}.  In that case, writing the horizon generating Killing field in the Fefferman-Graham coordinates of \cite{Fischetti:2012ps} leads to a negative $t$ component on part of the horizon, even though this component is everywhere positive at the AlAdS boundary.  It would also be interesting to transform our current 3+1 solutions to the coordinates of \cite{Bhattacharyya:2008xc} (say, for a solution deep within the hydrodynamic regime), though the additional numerics required places such an analysis is beyond the scope of this work.

\begin{figure}[t]
\centering
\subfigure[]{
    \includegraphics[width=0.3\textwidth]{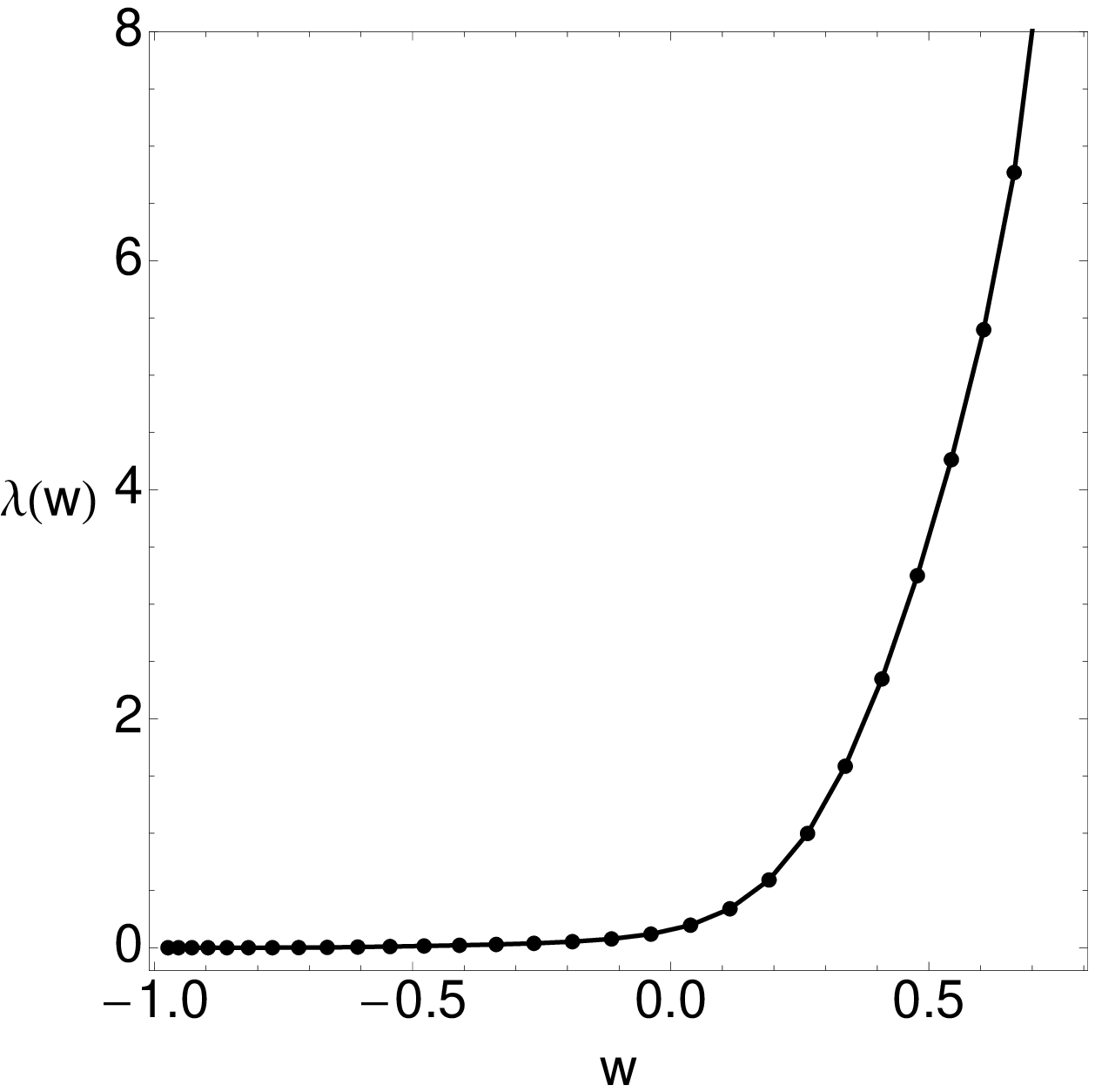}
    \label{fig:affinehorizon}
}
\subfigure[]{
    \includegraphics[width=0.32\textwidth]{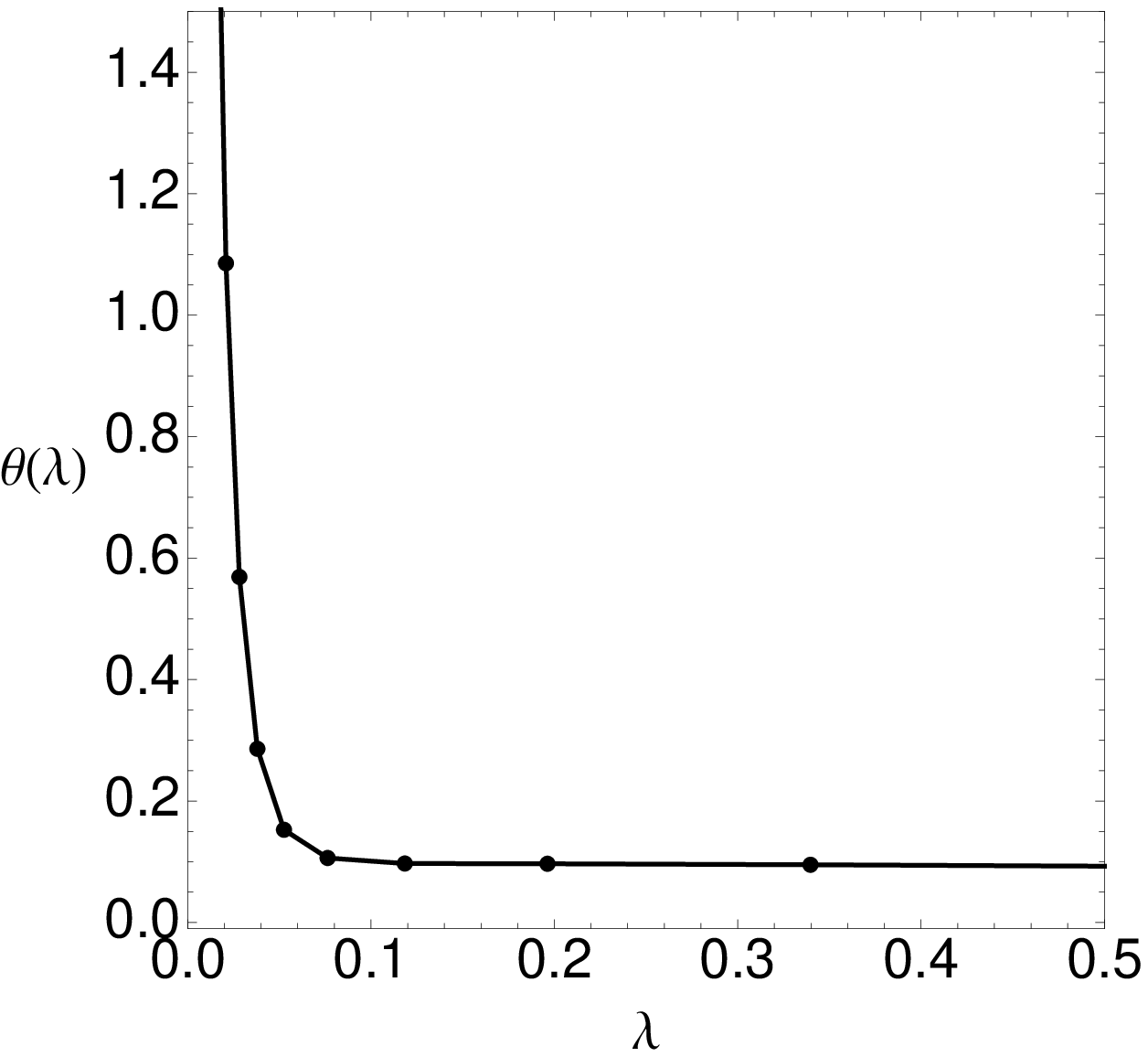}
    \label{fig:expansionhorizon}
}
\subfigure[]{
    \includegraphics[width=0.31\textwidth]{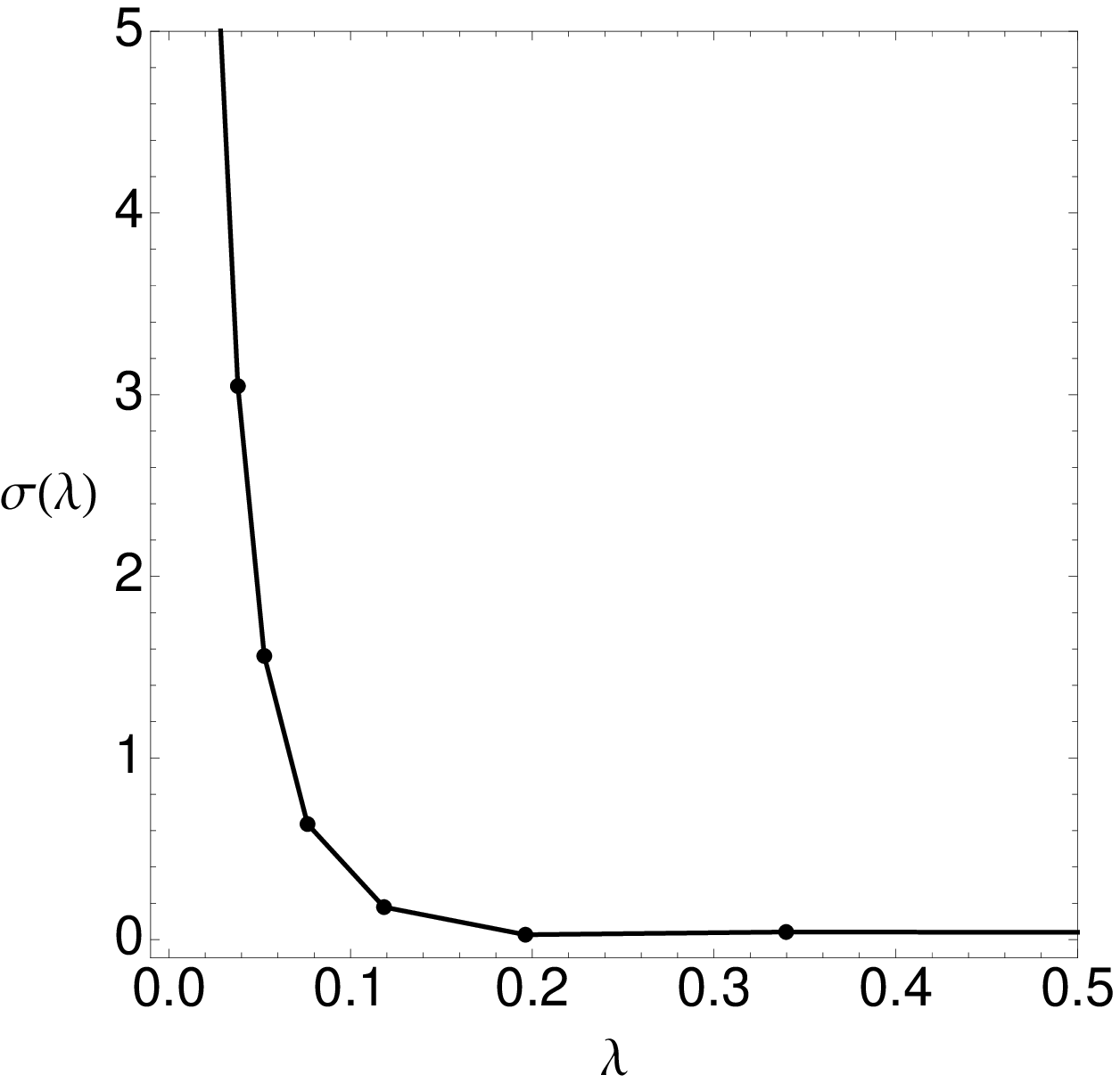}
    \label{fig:shear}
}
\caption[Optional caption for list of figures]{(a): An affine parameter along the horizon as a function of $w$.  (b): The expansion of a future horizon generator as a function of $\lambda$.  (c): The positive eigenvalue $\sigma$ of ${\hat{\sigma}^I}_{\ J}$ as a function of $\lambda$.  At small $\lambda$ we find $\sigma \sim \lambda^{-5/6}$.  All figures use $\alpha_\pm = 1$ and $\Delta T/T_{\infty} = 0.2$.}
\label{figs:horizon}
\end{figure}

We may now proceed to investigate various quantities along the horizon. Perhaps the most important quantity is the affine parameter $\lambda$, which we show in  Fig.~\ref{fig:affinehorizon} as function of $w$.  Note that $\lambda$ approaches a constant value at $w=-1$. This is to be expected, as we have already noted that $w=-1$ is the past horizon.  Since the affine parameter is only defined up to affine transformations, this constant is arbitrary and we have set $\lambda(w=-1) =0$ for convenience.  In contrast, the affine parameter diverges as we approach $w=1$.

Figure~\ref{fig:expansionhorizon} shows the expansion $\theta$ as a function of $\lambda$.  As expected on general grounds, $\theta$ is everywhere positive with $d\theta /d\lambda<0$ and $\theta$ asymptotes to zero at large $\lambda$. We see this as the most solid test of our numerics. Note that the sign of $d\theta /d\lambda<0$ is only guaranteed via Raychaudhuri's equation once the equations of motion are used. It is thus far from trivial that the sign comes out right.

The expansion diverges at the past horizon $(\lambda =0)$, indicating the presence of a caustic. In fact, it is easy to see that this caustic is a curvature singularity.  To do so, note from \ref{fig:hcomponents} that   $|\partial_\phi|$ diverges on the past horizon.  But since Killing fields obey a second order differential equation governed by the Riemann tensor (see e.g. (C.3.6) of \cite{Wald:1984rg}) they can diverge at finite affine parameter only if $R_{abcd}$ diverges in all orthonormal frames.

We now turn to the shear tensor $\hat \sigma_{IJ}$. From \eqref{eq:Bdef}, \eqref{eq:sigma} we see that since $h_{IJ}$ is diagonal, the same is true of $\hat \sigma_{IJ}$ . Since $\hat \sigma_{IJ}$ is also symmetric and traceless, it is completely characterized by the positive eigenvalue $\sigma$ of ${\hat{\sigma}^I}_{\ J}$, where the index was raised with the inverse $h^{IJ}$ of $h_{IJ}$.  Note that $\sigma$ is a spacetime scalar.

This eigenvalue is plotted as a function of $\lambda$ in figure \ref{fig:shear}.
As one might expect, it diverges at the caustic.  What is interesting is that we find the same divergence structure for all $\alpha, \Delta T$ that we have studied.   We quantify this behavior by fitting $\sigma(\lambda)$ to a power law $\mu \, \lambda^{\eta}$ near $\lambda=0$.  We have extracted $\eta$ for about $400$ flowing funnels spanning the domain $(\alpha,\Delta T/T_{\infty})\in(1,0.7)\times(0,0.4)$. In all cases we find
$\eta = -0.82 \pm 0.03$, where this error is in fact the maximum deviation. We note that this number is remarkably close to $-5/6$.  We can then use the Raychaudhuri equation \eqref{eq:RC} and the standard evolution equation for the shear (see (F.34) of \cite{Carroll:2004st}) to again show that $R_{abcd}$ diverges on the past horizon.  In fact, for $\eta = -5/6$ one may show that some Weyl tensor component $C_{abcd} k^b k^d$ (where $k^a$ is an affinely parametrized tangent to a null generator of the horizon) must diverge like $\lambda^{-11/6}$.  This fact merits an analytic explanation which we are unable to offer at this time.

\section{Discussion}
\label{sec:disc}

The above work constructs `flowing funnel' stationary black hole solutions.  Such solutions describe heat flow between reservoirs at unequal temperatures $T_\pm$.  The particular solutions constructed are global AdS${}_4$ flowing funnels which may be thought of as deformations of the BTZ black string \eqref{eq:BTZstring}.   Thus each heat reservoir lies just outside a boundary black hole of temperature $\alpha_\pm T_\pm$.  For the case $\alpha_\pm =1$, the CFT${}_3$ duals of our bulk solutions describe heat transfer between two non-dynamical 3-dimensional black holes due to CFT${}_3$ Hawking radiation .

Our solutions display many properties expected on general grounds.  There is a connected ergoregion near the horizon where $\partial_t$ becomes spacelike.  In fact, all Killing fields are spacelike at the future horizon ${\cal H}$, so that ${\cal H}$ is not a Killing horizon. This is consistent with the rigidity theorems \cite{R1,R2,Hollands:2006rj} since ${\cal H}$ is not compact.

The expansion $\theta$ of the null generators is everywhere positive but decreases toward the future along each null generator. The generators extend to infinite affine parameter in the far future (where $\theta \rightarrow 0$) but reach a caustic ($\theta \rightarrow \infty$) at finite affine parameter toward the past.   This caustic occurs on the past horizon, which is a curvature singularity characterized by a universal power law divergences for the shear $\sigma \sim \lambda^{-5/6}$ and for certain components of the Weyl tensor which grow like $\lambda^{-11/6}$ in any orthonormal frame.  These exponents were found numerically, but merit an analytic understanding.   It remains an open question whether curvature scalars (e.g. the Kretschmann scalar $R_{abcd} R^{abcd}$) might remain finite\footnote{Due to the large number of terms involved, we were not able to reliably calculate $R_{abcd} R^{abcd}$ from our numerics.}.  It would also be interesting to study the exponents governing the divergence of the expansion $\theta$, the norm $|\partial_\phi|^2$ and the inverse norm $|\partial_t|^{-2}$, though these have proved to be more difficult to extract from our numerics.

Note that $|\partial_\phi|$ decreases with $\lambda$ along the early part of the future horizon.  But since $\theta > 0$, the shrinkage of the $\phi$ circle with affine parameter $\lambda$ is more than compensated by the growth in $|\partial_t|$.  This positive expansion is associated with the expected generation of entropy due to the transport of heat from a hot source to a cold sink.  In particular, it is the analogue at large $\alpha_\pm, \Delta T/T_\infty$ of the entropy generation term \eqref{eq:sgen} seen in the hydrodynamic approximation.

In our coordinate system, the horizon generators appear to flow toward the hotter black hole.  While we have confirmed that this is a coordinate artifact, it would nevertheless be desirable to understand the effect in more detail.

We studied the boundary stress tensor of such solutions both numerically and to first order in the hydrodynamic (fluid/gravity) approximation. In particular, we computed the total heat flux $\Phi$ for boundary metrics of the form \eqref{eq:BBH} as a function of $\alpha, \Delta T/T_\infty$; see figure \ref{fig:boundaryflux3d}. It would clearly be of interest to study more general boundary metrics to understand which parts of this function are universal and which depend on detailed features of the boundary metric.

The hydrodynamic approximation is a derivative expansion which, since we fix all other parameters in the boundary metric, is for us governed by the parameters $\alpha_\pm$ (which control the extent to which the bulk and boundary black hole temperatures are detuned) and $\Delta T/(T_+ + T_-)$ (which controls the temperature difference between the heat source and sink).  As expected, we find excellent quantitative agreement when these parameters are small.  Interestingly, we also find good qualitative agreement when these parameters are close to $1$.  This gives yet another confirmation of the robust nature of the fluid/gravity correspondence as seen previously in e.g. \cite{Heller:2011ju}.
Of course, at large enough values of $\alpha_\pm, \Delta T/(T_+ + T_-)$ we find both quantitative and qualitative disagreement.   It would be interesting  see to what extent agreement might be improved by incorporating higher order hydrodynamic corrections. A particularly notable feature at large $\alpha$ ($\alpha_+ \gtrsim 1$ in our simulations) is that, while $T^t{}_{t}$ is always negative in the hydrodynamic limit, it becomes positive close to the hotter boundary black hole black hole. It would be interesting to understand this feature analytically.

\subsection*{Acknowledgements}
We thank Allen Adams, Pau Figueras, Veronika Hubeny,  Mukund Rangamani, and Toby Wiseman for many stimulating discussions of flowing funnels and methods by which they might be constructed.  We also thank Gary Horowitz for spotting an error in an early version of this manuscript.  This work was supported in part by the National Science Foundation under Grant Nos PHY11-25915
and PHY08-55415, and by funds from the University of California. DM also thanks the University of Colorado, Boulder, for its hospitality during the completion of this work.

\appendix

\section{Fluid results in the black hole frame}
\label{app:bhf}

We may transform the hydrodynamic results of section \ref{sec:firstorder} to the black hole frame associated with the metric \eqref{eq:BBH} by implementing a boundary conformal transformation and an appropriate change of coordinates.  Setting $\ell_4=1$ the result is
\begin{align}
\T_\mathrm{loc} &= \frac{y_0}{(1-\rho^2)G(\rho)} \left[\T_\infty + \frac{\Delta \T}{I} f(\rho) + \mathcal{O}(\Delta \T^2)\right], \\
\label{eq:linBHF1}
16\pi G^{(4)} {T^t}_t &= \frac{y_0^3}{(1-\rho^2)^3 G^3(\rho)} \left[ -2\T_\infty^3 - \frac{6\T_\infty^2 \Delta \T}{I} \, f(\rho) + \mathcal{O}(\Delta \T^2)\right], \\
16\pi G^{(4)} {T^t}_\rho &= -\frac{9\T_\infty^3 \Delta \T}{I} \, \frac{y_0^3}{\sqrt{2-\rho^2} \, (1-\rho^2)^3G^3(\rho)} + \mathcal{O}(\Delta \T^2), \\
16\pi G^{(4)} {T^\rho}_\rho &= \frac{y_0^3}{(1-\rho^2)^3 G^3(\rho)} \left[\T_\infty^3 + \frac{3\T_\infty^2 \Delta \T}{I}\, \left(h(\rho) + f(\rho) \right) + \mathcal{O}(\Delta \T^2)\right], \\
16\pi G^{(4)} {T^\phi}_\phi &= \frac{y_0^3}{(1-\rho^2)^3 G^3(\rho)} \left[\T_\infty^3 + \frac{3\T_\infty^2 \Delta \T}{I}\, \left(- h(\rho) + f(\rho) \right) + \mathcal{O}(\Delta \T^2)\right],
\label{eq:linBHF4}
\end{align}
where $\Delta \T$ is again defined with respect to $\partial_t$, $\T_{\mathrm{loc}}$ is the local temperature with respect to proper time in the fluid rest frame, and setting~$H(\rho) := (1-\rho^2) G(\rho)/y_0$ we have defined
\begin{align}
h(\rho) &= \frac{1}{2} \sqrt{2-\rho^2} \, H(\rho) H'(\rho), \\
f(\rho) &= \int_0^\rho \frac{1}{2\sqrt{2-\rho^2}} \left[\rho H(\rho) H'(\rho) + (2-\rho^2) \left(\left(H'(\rho)\right)^2 - H(\rho) H''(\rho)\right)\right] \, d\rho, \\
I &= f(1) - f(-1).
\end{align}

\section{The horizon-generating null congruence}
\label{sec:horizon}

We wish to study the expansion and the shear tensor associated with the null geodesic congruence that generates the future bulk horizon.  Instead of solving the geodesic equation and taking derivatives of deviation vectors, we take advantage of the fact that our system is co-homogeneity 2 to compute these quantities directly (up to a position-dependent scale factor) from the induced metric $h_{IJ}$ on 2-dimensional surfaces tangent to the Killing fields $\partial_t, \partial_\phi$.  We then compute the affine parameter $\lambda$ along these geodesics from the Raychaudhuri equation as explained below.

Recall that we consider a future event horizon ${\cal H}$ of an AlAdS${}_4$ spacetime with two commuting KVFs $\partial_t$ and $\partial_\phi$.  The horizon is 3-dimensional, with a two-dimensional space of generators.  So long as the horizon is not itself Killing, we see that any two generators are related by the actions of $\partial_t$ and $\partial_\phi$.

Choose one horizon generator with affine parameter $\lambda$.  We can extend $\lambda$ to a scalar function on ${\cal H}$ by requiring it to be invariant under $\partial_t, \partial_\phi$.  In our case we can take $\lambda = \lambda(w)$ since $w$ is indeed invariant under both KVFs and is a good coordinate on ${\cal H}$.

Let $k^a$ be the tangent to our generator associated with affine parameter $\lambda$.  Note that since $\partial_t, \partial_\phi$ are also tangent to the horizon we have $k \perp \partial_t, \partial_\phi$.  We also choose any $\ell^a$ satisfying $\ell^a k_a =-1$ and $\ell \perp \partial_t, \partial_\phi$.  We then extend $k,\ell$ to vector fields defined across all of ${\cal H}$ by requiring them to be invariant under $\partial_t, \partial_\phi$.  We then define a `deformation tensor' $\hat B_{ab}$ associated with flow along the horizon generators by projecting $B_{ab} = \nabla_b k_a$ onto the space orthogonal to $k,\ell$.  See e.g. appendix F of \cite{Carroll:2004st}.

Let us note that since $\partial_t, \partial_\phi$ commute they are surface-forming, and $k$ is orthogonal to this surface.  So $k$ is hypersurface orthogonal and the twist $\hat \omega_{ab} = \hat B_{[ab]}$ vanishes.  Thus $\hat B_{ab} = \hat B_{ba}$.  We will use this symmetry below.

Deviation vector fields for the horizon-generating null congruence are defined by the property that, when evaluated on a given horizon generator $\gamma$, they point to the same horizon generator $\gamma'$ for all $\lambda$. Let us consider a deviation vector field $\eta$ orthogonal to both $k$ and $\ell$.
Then (see e.g. appendix F of \cite{Carroll:2004st}) $\eta$ satisfies
\begin{equation}
\label{eq:etaE}
\eta^c \hat B^a_c = k^c \nabla_c \eta^a .
\end{equation}

Since translations along $\partial_t, \partial_\phi$ map one geodesic to another, both  $\partial_t$ and  $\partial_\phi$ are deviation vectors.  And both are orthogonal to $k, \ell$.  So we may choose $\eta_I = \partial_I$ for $I=t,\phi$.  Here the $\eta_I$ are two spacetime vectors, not the components of a single vector.

Let us now consider the set of associated inner products
\begin{equation}
h_{IJ} = \eta_I \cdot \eta_J := \eta_I^a g_{ab} \eta_J^b .
\end{equation}
In any coordinate system,
$\eta_t^a = \partial_t x^a$ and $\eta_\phi^a = \partial_\phi x^a$.  So in particular in the coordinate system $y,w,\tilde {t},\phi$ we have (since $\partial_{\tilde t} = \partial_t$)
\begin{equation}
\label{eq:ind}
h_{IJ} = g_{IJ};
\end{equation}
i.e., this is just the induced metric on the 2-plane generated by $\partial_t, \partial_\phi$ in coordinates $(\tilde t,\phi)$ or, equivalently for this purpose, coordinates $(t,\phi)$.  So it is easy to read off from our numerics.  But note that $h_{IJ}$ was defined to be a set of scalars, so covariant derivatives of $h_{IJ}$ are just coordinate derivatives.

The evolution of $h_{IJ}$ (with respect to $\lambda$, or equivalently with respect to $w$) is governed by \eqref{eq:etaE}. Using \eqref{eq:etaE} we compute

\begin{eqnarray}\label{eq:Bdef}
\frac{d}{d\lambda} h_{IJ} &= & k^c \nabla_c h_{IJ} = k^c \nabla_c \left( \eta_I \cdot \eta_J \right) \cr
&= &  \hat B^a_c \eta^c_I \eta_{Ja} +  \eta^a_I \hat B^c_a  \eta_{Jc} \cr
&= &  2 \hat B_{ac} \eta^a_I \eta^c_{J} = 2 \hat B_{IJ},
\end{eqnarray}
where in the last step as for \eqref{eq:ind} above we have used $\hat B_{IJ}$ to denote the $\tilde {t}, \phi$ components of $\hat B_{ac}$ in the particular coordinate system $y,w,\tilde {t},\phi$ (or equivalently the $t,\phi$ components).

The deformation tensor $\hat B^{ab}$ is by construction orthogonal to $k,\ell$.  Thus we may write $\hat B^{ab} \partial_a \partial_b = \hat B^{IJ} \partial_I \partial_J$.   Furthermore, from \eqref{eq:ind} we have
\begin{equation}
\hat B_{IJ} = g_{Ia}\hat B^{ab}  g_{Jb} = g_{IK}\hat B^{KL}  g_{LJ}
=h_{IK}\hat B^{KL}  h_{LJ},
\end{equation}
where $K,L$ also range over $\phi,t$.  Thus we may safely use $h_{IJ}$ and its inverse $h^{IJ}$ to raise and lower indices $I,J$ on $\hat B_{IJ}$.

Now, the components $\hat B_{IJ}$ are essentially exponentials of integrated versions of the expansion and shear.  In particular, introducing the projector ${Q^a}_b$ onto the subspace orthogonal to $k,\ell$ (i.e., onto the space spanned by $\partial_t, \partial_\phi$) we have

\begin{equation}
\label{eq:theta}
\theta = Q_{ab}  \hat B^{ab} = h_{IJ} \hat B^{IJ} = h^{IJ} \hat B_{IJ},
\end{equation}
and
\begin{equation}
\label{eq:sigma}
\hat \sigma_{IJ} = \hat B_{IJ} - \frac{1}{D-2}\theta h_{IJ},
\end{equation}
where $D=4$ for AdS${}_4$.

Of course, it remains to actually find the affine parameter $\lambda$ used in the above definitions.  We choose to calculate $\lambda$ from Raychaudhuri's equation which, in the present context, may be written
\begin{equation}
\label{eq:RC}
\frac{d \theta}{d\lambda} = - \hat B_{ab} \hat B^{ab} - Q^{ab} R_{acbd} k^c k^d .
\end{equation}
As usual, the symmetries of the Riemann tensor imply that $Q^{ab} R_{acbd} k^c k^d = g^{ab} R_{acbd} k^c k^d$, which is proportional to $R_{cd} k^c k^d$.  But $R_{ab} \propto g_{ab}$ by the equations of motion, so the final term in \eqref{eq:RC} vanishes.  Since $\hat B^{ab}$ is orthogonal to both $k$ and $\ell$ we may then write
\begin{equation}
\label{eq:RC2}
\frac{d \theta}{d\lambda} = - \hat B_{IJ} \hat B^{IJ}.
\end{equation}
In terms of a general coordinate $w$ along the generators, \eqref{eq:RC2} may be rearranged to yield
\begin{equation}
\lambda'' = \lambda' \left(h^{IJ} h'_{IJ}\right)^{-1} \left[  \frac{1}{2} h^{I_1 I_2} h^{J_1 J_2} h'{}_{I_1 J_1} h'{}_{I_2 J_2}  + \frac{d}{dw} \left( h^{IJ} h'{}_{IJ} \right) \right] := \lambda' Z(w),
\end{equation}
where $'$ denotes the coordinate derivative $d/dw$ and the last equality defines $Z(w)$.  This equation is then easily solved for $\lambda$ in terms $Z(w)$, which is relatively straightforward to extract from the numerics.


\bibliographystyle{JHEP}
\bibliography{Funnelsbib}

\providecommand{\href}[2]{#2}\begingroup\raggedright\begin{thebibliography}{10}

\bibitem{Son:2006em}
D.~T. Son and A.~O. Starinets, {\it {Hydrodynamics of r-charged black holes}},
  {\em JHEP} {\bf 0603} (2006) 052,
  [\href{http://xxx.lanl.gov/abs/hep-th/0601157}{{\tt hep-th/0601157}}].

\bibitem{Chesler:2010bi}
P.~M. Chesler and L.~G. Yaffe, {\it {Holography and colliding gravitational
  shock waves in asymptotically AdS5 spacetime}},  {\em Phys.Rev.Lett.} {\bf
  106} (2011) 021601, [\href{http://xxx.lanl.gov/abs/1011.3562}{{\tt
  arXiv:1011.3562}}].

\bibitem{Heller:2011ju}
M.~P. Heller, R.~A. Janik, and P.~Witaszczyk, {\it {The characteristics of
  thermalization of boost-invariant plasma from holography}},  {\em
  Phys.Rev.Lett.} {\bf 108} (2012) 201602,
  [\href{http://xxx.lanl.gov/abs/1103.3452}{{\tt arXiv:1103.3452}}].

\bibitem{Bantilan:2012vu}
H.~Bantilan, F.~Pretorius, and S.~S. Gubser, {\it {Simulation of Asymptotically
  AdS5 Spacetimes with a Generalized Harmonic Evolution Scheme}},  {\em
  Phys.Rev.} {\bf D85} (2012) 084038,
  [\href{http://xxx.lanl.gov/abs/1201.2132}{{\tt arXiv:1201.2132}}].

\bibitem{Khlebnikov:2010yt}
S.~Khlebnikov, M.~Kruczenski, and G.~Michalogiorgakis, {\it {Shock waves in
  strongly coupled plasmas}},  {\em Phys.Rev.} {\bf D82} (2010) 125003,
  [\href{http://xxx.lanl.gov/abs/1004.3803}{{\tt arXiv:1004.3803}}].

\bibitem{Khlebnikov:2011ka}
S.~Khlebnikov, M.~Kruczenski, and G.~Michalogiorgakis, {\it {Shock waves in
  strongly coupled plasmas II}},  {\em JHEP} {\bf 1107} (2011) 097,
  [\href{http://xxx.lanl.gov/abs/1105.1355}{{\tt arXiv:1105.1355}}].

\bibitem{Maldacena:1997re}
J.~M. Maldacena, {\it {The Large N limit of superconformal field theories and
  supergravity}},  {\em Adv.Theor.Math.Phys.} {\bf 2} (1998) 231--252,
  [\href{http://xxx.lanl.gov/abs/hep-th/9711200}{{\tt hep-th/9711200}}].

\bibitem{AM}
D.~Astefanesei and R.~C. Myers, ``{Boundary black holes and ads/cft
  correspondence}.''
\newblock Talk presented by R.C. Myers at Black Holes IV: Theory and
  Mathematical Aspects, at Honey Harbor, Ontario, May 25-28, 2003.

\bibitem{Tanaka:2002rb}
T.~Tanaka, {\it {Classical black hole evaporation in Randall-Sundrum infinite
  brane world}},  {\em Prog.Theor.Phys.Suppl.} {\bf 148} (2003) 307--316,
  [\href{http://xxx.lanl.gov/abs/gr-qc/0203082}{{\tt gr-qc/0203082}}].

\bibitem{Emparan:2002px}
R.~Emparan, A.~Fabbri, and N.~Kaloper, {\it {Quantum black holes as holograms
  in AdS brane worlds}},  {\em JHEP} {\bf 0208} (2002) 043,
  [\href{http://xxx.lanl.gov/abs/hep-th/0206155}{{\tt hep-th/0206155}}].

\bibitem{Wiseman:2001xt}
T.~Wiseman, {\it {Relativistic stars in Randall-Sundrum gravity}},  {\em
  Phys.Rev.} {\bf D65} (2002) 124007,
  [\href{http://xxx.lanl.gov/abs/hep-th/0111057}{{\tt hep-th/0111057}}].

\bibitem{Wiseman:2002zc}
T.~Wiseman, {\it {Static axisymmetric vacuum solutions and nonuniform black
  strings}},  {\em Class.Quant.Grav.} {\bf 20} (2003) 1137--1176,
  [\href{http://xxx.lanl.gov/abs/hep-th/0209051}{{\tt hep-th/0209051}}].

\bibitem{Casadio:2002uv}
R.~Casadio and L.~Mazzacurati, {\it {Bulk shape of brane world black holes}},
  {\em Mod.Phys.Lett.} {\bf A18} (2003) 651--660,
  [\href{http://xxx.lanl.gov/abs/gr-qc/0205129}{{\tt gr-qc/0205129}}].

\bibitem{Karasik:2003tx}
D.~Karasik, C.~Sahabandu, P.~Suranyi, and L.~Wijewardhana, {\it {Small black
  holes in Randall-Sundrum I scenario}},  {\em Phys.Rev.} {\bf D69} (2004)
  064022, [\href{http://xxx.lanl.gov/abs/gr-qc/0309076}{{\tt gr-qc/0309076}}].

\bibitem{Kudoh:2003xz}
H.~Kudoh, T.~Tanaka, and T.~Nakamura, {\it {Small localized black holes in
  brane world: Formulation and numerical method}},  {\em Phys.Rev.} {\bf D68}
  (2003) 024035, [\href{http://xxx.lanl.gov/abs/gr-qc/0301089}{{\tt
  gr-qc/0301089}}].

\bibitem{Kudoh:2003vg}
H.~Kudoh, {\it {Thermodynamical properties of small localized black hole}},
  {\em Prog.Theor.Phys.} {\bf 110} (2004) 1059--1069,
  [\href{http://xxx.lanl.gov/abs/hep-th/0306067}{{\tt hep-th/0306067}}].

\bibitem{Kudoh:2004kf}
H.~Kudoh, {\it {Six-dimensional localized black holes: Numerical solutions}},
  {\em Phys.Rev.} {\bf D69} (2004) 104019,
  [\href{http://xxx.lanl.gov/abs/hep-th/0401229}{{\tt hep-th/0401229}}].

\bibitem{Karasik:2004wk}
D.~Karasik, C.~Sahabandu, P.~Suranyi, and L.~Wijewardhana, {\it {Small black
  holes on branes: Is the horizon regular or singular?}},  {\em Phys.Rev.} {\bf
  D70} (2004) 064007, [\href{http://xxx.lanl.gov/abs/gr-qc/0404015}{{\tt
  gr-qc/0404015}}].

\bibitem{Fitzpatrick:2006cd}
A.~L. Fitzpatrick, L.~Randall, and T.~Wiseman, {\it {On the existence and
  dynamics of braneworld black holes}},  {\em JHEP} {\bf 0611} (2006) 033,
  [\href{http://xxx.lanl.gov/abs/hep-th/0608208}{{\tt hep-th/0608208}}].

\bibitem{Yoshino:2008rx}
H.~Yoshino, {\it {On the existence of a static black hole on a brane}},  {\em
  JHEP} {\bf 0901} (2009) 068, [\href{http://xxx.lanl.gov/abs/0812.0465}{{\tt
  arXiv:0812.0465}}].

\bibitem{Gregory:2008br}
R.~Gregory, S.~F. Ross, and R.~Zegers, {\it {Classical and quantum gravity of
  brane black holes}},  {\em JHEP} {\bf 0809} (2008) 029,
  [\href{http://xxx.lanl.gov/abs/0802.2037}{{\tt arXiv:0802.2037}}].

\bibitem{Hubeny:2009ru}
V.~E. Hubeny, D.~Marolf, and M.~Rangamani, {\it {Hawking radiation in large N
  strongly-coupled field theories}},  {\em Class.Quant.Grav.} {\bf 27} (2010)
  095015, [\href{http://xxx.lanl.gov/abs/0908.2270}{{\tt arXiv:0908.2270}}].

\bibitem{Hubeny:2009kz}
V.~E. Hubeny, D.~Marolf, and M.~Rangamani, {\it {Black funnels and droplets
  from the AdS C-metrics}},  {\em Class.Quant.Grav.} {\bf 27} (2010) 025001,
  [\href{http://xxx.lanl.gov/abs/0909.0005}{{\tt arXiv:0909.0005}}].

\bibitem{Hubeny:2009rc}
V.~E. Hubeny, D.~Marolf, and M.~Rangamani, {\it {Hawking radiation from AdS
  black holes}},  {\em Class.Quant.Grav.} {\bf 27} (2010) 095018,
  [\href{http://xxx.lanl.gov/abs/0911.4144}{{\tt arXiv:0911.4144}}].

\bibitem{Caldarelli:2011wa}
M.~M. Caldarelli, O.~J. Dias, R.~Monteiro, and J.~E. Santos, {\it {Black
  funnels and droplets in thermal equilibrium}},  {\em JHEP} {\bf 1105} (2011)
  116, [\href{http://xxx.lanl.gov/abs/1102.4337}{{\tt arXiv:1102.4337}}].
  Temporary entry.

\bibitem{Kleihaus:2011yq}
B.~Kleihaus, J.~Kunz, E.~Radu, and D.~Senkbeil, {\it {Electric charge on the
  brane?}},  {\em Phys.Rev.} {\bf D83} (2011) 104050,
  [\href{http://xxx.lanl.gov/abs/1103.4758}{{\tt arXiv:1103.4758}}].

\bibitem{Headrick:2009pv}
M.~Headrick, S.~Kitchen, and T.~Wiseman, {\it {A New approach to static
  numerical relativity, and its application to Kaluza-Klein black holes}},
  {\em Class.Quant.Grav.} {\bf 27} (2010) 035002,
  [\href{http://xxx.lanl.gov/abs/0905.1822}{{\tt arXiv:0905.1822}}].

\bibitem{Figueras:2011va}
P.~Figueras, J.~Lucietti, and T.~Wiseman, {\it {Ricci solitons, Ricci flow, and
  strongly coupled CFT in the Schwarzschild Unruh or Boulware vacua}},  {\em
  Class.Quant.Grav.} {\bf 28} (2011) 215018,
  [\href{http://xxx.lanl.gov/abs/1104.4489}{{\tt arXiv:1104.4489}}]. Temporary
  entry.

\bibitem{Figueras:2011gd}
P.~Figueras and T.~Wiseman, {\it {Gravity and large black holes in
  Randall-Sundrum II braneworlds}},  {\em Phys.Rev.Lett.} {\bf 107} (2011)
  081101, [\href{http://xxx.lanl.gov/abs/1105.2558}{{\tt arXiv:1105.2558}}].
  Temporary entry.

\bibitem{Fischetti:2012ps}
S.~Fischetti and D.~Marolf, {\it {Flowing Funnels: Heat sources for field
  theories and the AdS3 dual of CFT2 Hawking radiation}},  {\em
  Class.Quant.Grav.} {\bf 29} (2012) 105004,
  [\href{http://xxx.lanl.gov/abs/1202.5069}{{\tt arXiv:1202.5069}}].

\bibitem{Santos:2012he}
J.~E. Santos and B.~Way, {\it {Black Funnels}},  {\em JHEP} {\bf 1212} (2012)
  060, [\href{http://xxx.lanl.gov/abs/1208.6291}{{\tt arXiv:1208.6291}}].

\bibitem{Banados:1992wn}
M.~Banados, C.~Teitelboim, and J.~Zanelli, {\it {The Black hole in
  three-dimensional space-time}},  {\em Phys.Rev.Lett.} {\bf 69} (1992)
  1849--1851, [\href{http://xxx.lanl.gov/abs/hep-th/9204099}{{\tt
  hep-th/9204099}}].

\bibitem{Banados:1992gq}
M.~Banados, M.~Henneaux, C.~Teitelboim, and J.~Zanelli, {\it {Geometry of the
  (2+1) black hole}},  {\em Phys.Rev.} {\bf D48} (1993) 1506--1525,
  [\href{http://xxx.lanl.gov/abs/gr-qc/9302012}{{\tt gr-qc/9302012}}].

\bibitem{Bhattacharyya:2008jc}
S.~Bhattacharyya, V.~E. Hubeny, S.~Minwalla, and M.~Rangamani, {\it {Nonlinear
  Fluid Dynamics from Gravity}},  {\em JHEP} {\bf 0802} (2008) 045,
  [\href{http://xxx.lanl.gov/abs/0712.2456}{{\tt arXiv:0712.2456}}].

\bibitem{Kinoshita:2011qs}
S.~Kinoshita and N.~Tanahashi, {\it {Hawking temperature for near-equilibrium
  black holes}},  {\em Phys.Rev.} {\bf D85} (2012) 024050,
  [\href{http://xxx.lanl.gov/abs/1111.2684}{{\tt arXiv:1111.2684}}].

\bibitem{Abreu:2010ru}
G.~Abreu and M.~Visser, {\it {Kodama time: Geometrically preferred foliations
  of spherically symmetric spacetimes}},  {\em Phys.Rev.} {\bf D82} (2010)
  044027, [\href{http://xxx.lanl.gov/abs/1004.1456}{{\tt arXiv:1004.1456}}].

\bibitem{R1}
S.~W. Hawking, {\it {Black holes in general relativity}},  {\em Commun. Math.
  Phys.} {\bf 25} (1972) 152--166.

\bibitem{R2}
S.~W. Hawking and G.~F.~R. Ellis, {\em {The large scale structure of
  space-time}}.
\newblock Cambridge University Press, Cambridge, UK, 1973.

\bibitem{Hollands:2006rj}
S.~Hollands, A.~Ishibashi, and R.~M. Wald, {\it {A Higher dimensional
  stationary rotating black hole must be axisymmetric}},  {\em
  Commun.Math.Phys.} {\bf 271} (2007) 699--722,
  [\href{http://xxx.lanl.gov/abs/gr-qc/0605106}{{\tt gr-qc/0605106}}].

\bibitem{Emparan:1999fd}
R.~Emparan, G.~T. Horowitz, and R.~C. Myers, {\it {Exact description of black
  holes on branes. 2. Comparison with BTZ black holes and black strings}},
  {\em JHEP} {\bf 0001} (2000) 021,
  [\href{http://xxx.lanl.gov/abs/hep-th/9912135}{{\tt hep-th/9912135}}].

\bibitem{FW}
P.~Figueras and T.~Wiseman, ``{Stationary holographic plasma quenches and
  numerical methods for non-Killing horizons}.''.

\bibitem{Headrick:2010zt}
M.~Headrick, {\it {Entanglement Renyi entropies in holographic theories}},
  {\em Phys.Rev.} {\bf D82} (2010) 126010,
  [\href{http://xxx.lanl.gov/abs/1006.0047}{{\tt arXiv:1006.0047}}].

\bibitem{Marolf:2010tg}
D.~Marolf, M.~Rangamani, and M.~Van~Raamsdonk, {\it {Holographic models of de
  Sitter QFTs}},  {\em Class.Quant.Grav.} {\bf 28} (2011) 105015,
  [\href{http://xxx.lanl.gov/abs/1007.3996}{{\tt arXiv:1007.3996}}].

\bibitem{Hung:2011nu}
L.-Y. Hung, R.~C. Myers, M.~Smolkin, and A.~Yale, {\it {Holographic
  Calculations of Renyi Entropy}},  {\em JHEP} {\bf 1112} (2011) 047,
  [\href{http://xxx.lanl.gov/abs/1110.1084}{{\tt arXiv:1110.1084}}].

\bibitem{Emparan:1998he}
R.~Emparan, {\it {AdS membranes wrapped on surfaces of arbitrary genus}},  {\em
  Phys.Lett.} {\bf B432} (1998) 74--82,
  [\href{http://xxx.lanl.gov/abs/hep-th/9804031}{{\tt hep-th/9804031}}].

\bibitem{Birmingham:1998nr}
D.~Birmingham, {\it {Topological black holes in Anti-de Sitter space}},  {\em
  Class.Quant.Grav.} {\bf 16} (1999) 1197--1205,
  [\href{http://xxx.lanl.gov/abs/hep-th/9808032}{{\tt hep-th/9808032}}].

\bibitem{Emparan:1999gf}
R.~Emparan, {\it {AdS / CFT duals of topological black holes and the entropy of
  zero energy states}},  {\em JHEP} {\bf 9906} (1999) 036,
  [\href{http://xxx.lanl.gov/abs/hep-th/9906040}{{\tt hep-th/9906040}}].

\bibitem{Rangamani:2009xk}
M.~Rangamani, {\it {Gravity and Hydrodynamics: Lectures on the fluid-gravity
  correspondence}},  {\em Class.Quant.Grav.} {\bf 26} (2009) 224003,
  [\href{http://xxx.lanl.gov/abs/0905.4352}{{\tt arXiv:0905.4352}}].

\bibitem{Das:2010mk}
S.~R. Das, A.~Ghosh, J.-H. Oh, and A.~D. Shapere, {\it {On Dumb Holes and their
  Gravity Duals}},  {\em JHEP} {\bf 1104} (2011) 030,
  [\href{http://xxx.lanl.gov/abs/1011.3822}{{\tt arXiv:1011.3822}}].

\bibitem{deHaro:2000xn}
S.~de~Haro, S.~N. Solodukhin, and K.~Skenderis, {\it {Holographic
  reconstruction of space-time and renormalization in the AdS / CFT
  correspondence}},  {\em Commun.Math.Phys.} {\bf 217} (2001) 595--622,
  [\href{http://xxx.lanl.gov/abs/hep-th/0002230}{{\tt hep-th/0002230}}].

\bibitem{Adam:2011dn}
A.~Adam, S.~Kitchen, and T.~Wiseman, {\it {A numerical approach to finding
  general stationary vacuum black holes}},  {\em Class.Quant.Grav.} {\bf 29}
  (2012) 165002, [\href{http://xxx.lanl.gov/abs/1105.6347}{{\tt
  arXiv:1105.6347}}].

\bibitem{Bhattacharyya:2008xc}
S.~Bhattacharyya, V.~E. Hubeny, R.~Loganayagam, G.~Mandal, S.~Minwalla, et~al.,
  {\it {Local Fluid Dynamical Entropy from Gravity}},  {\em JHEP} {\bf 0806}
  (2008) 055, [\href{http://xxx.lanl.gov/abs/0803.2526}{{\tt
  arXiv:0803.2526}}].

\bibitem{Wald:1984rg}
R.~M. Wald, {\em {General Relativity}}.
\newblock Chicago University Press, Chicago, USA, 1984.

\bibitem{Carroll:2004st}
S.~M. Carroll, {\em {Spacetime and geometry: An introduction to general
  relativity}}.
\newblock Addison-Wesley, San Francisco, USA, 2004.

\end{thebibliography}\endgroup

\end{document}